\documentclass[twocolumn,twocolappendix]{aastex63}
\usepackage{apjfonts}
\usepackage{graphicx}
\usepackage{multirow}
\usepackage{comment}
\usepackage{natbib}
\usepackage{hyperref}
\usepackage{amsmath}
\usepackage{fontenc}
\usepackage{verbatim}  
\usepackage[toc,page]{appendix} 
\usepackage{lineno}

\newcommand{\tnm}{\tablenotemark}

\newcommand{\lagn}{\mbox{$L_{5100}$}}
\newcommand{\mmue}{\mbox{$\langle \mu_{e,I} \rangle$}}

\def\hst{{HST}}
\def\galfit{{\tt GALFIT}}
\def\sersic{S\'{e}rsic}
\def\h{\hskip -1.5 mm}

\newcommand{\mbh}{\mbox{$M_{\rm BH}$}}
\newcommand{\mbulge}{\mbox{$M_{\rm bulge}$}}
\newcommand{\mhost}{\mbox{$M_{\rm total}$}}

\submitjournal{ApJ}

\shorttitle{Host Galaxies of Type 1 quasars}
\shortauthors{ZHAO ET AL.}

\begin{document}

\title{The Diverse Morphology, Stellar Population, and Black Hole Scaling Relations of the Host Galaxies of Nearby Quasars}

\correspondingauthor{Jinyi Shangguan}
\email{shangguan@mpe.mpg.de}

\author[0000-0003-4591-2532]{Yulin Zhao}
\affil{Kavli Institute for Astronomy and Astrophysics, Peking University, Beijing 100871, China}
\affiliation{Department of Astronomy, School of Physics, Peking University, Beijing 100871, China}

\author[0000-0001-6947-5846]{Luis C. Ho}
\affil{Kavli Institute for Astronomy and Astrophysics, Peking University, Beijing 100871, China}
\affiliation{Department of Astronomy, School of Physics, Peking University, Beijing 100871, China}

\author[0000-0002-4569-9009]{Jinyi Shangguan}
\affil{Max-Planck-Institut f\"{u}r Extraterrestrische Physik (MPE), Giessenbachstr., D-85748 Garching, Germany}
\affiliation{Kavli Institute for Astronomy and Astrophysics, Peking University, Beijing 100871, China}

\author[0000-0002-3560-0781]{Minjin Kim}
\affil{Department of Astronomy and Atmospheric Sciences, Kyungpook National University, Daegu 702-701, Republic of Korea}
\affiliation{Korea Astronomy and Space Science Institute, Daejeon 305-348, Republic of Korea}

\author[0000-0001-8592-7910]{Dongyao Zhao}
\affil{Beijing Planetarium, Beijing Academy of Science and Technology, Beijing, 100044, China}
\affiliation{Kavli Institute for Astronomy and Astrophysics, Peking University, Beijing 100871, China}

\author[0000-0003-1015-5367]{Hua Gao}
\affil{Kavli Institute for Astronomy and Astrophysics, Peking University, Beijing 100871, China}
\affiliation{Department of Astronomy, School of Physics, Peking University, Beijing 100871, China}
\affiliation{Kavli Institute for the Physics and Mathematics of the Universe (WPI),The University of Tokyo Institutes for Advanced Study,\\
The University of Tokyo, Kashiwa, Chiba 277-8583, Japan}

\begin{abstract}
We present rest-frame $B$ and $I$ imaging of 35 low-redshift ($z < 0.5$) Palomar-Green quasars using the Hubble Space Telescope Wide Field Camera 3.  We perform multi-component two-dimensional image decomposition to separate the host galaxy from its bright active nucleus, characterize its morphology, and measure its photometric properties.  Special care is devoted to quantifying the structural parameters of the galaxy bulge, determine its $B-I$ color, and estimate its stellar mass.  Roughly half of the sample, comprising the less luminous ($\lagn \lesssim 10^{45}\,\mathrm{erg\,s^{-1}}$) but most high Eddington ratio quasars, reside in disk galaxies that are often barred and possess pseudo bulges.  The large stellar masses, large effective radii, and faint surface brightnesses suggest that the host galaxies of the most luminous quasars are mostly ellipticals.  Major mergers constitute only a minority ($\lesssim 20\%$) of our sample.  Our quasar sample roughly obeys the scaling relations between black hole mass and host galaxy (bulge, core, total) stellar mass.  Hosts with black holes more massive than $\sim 10^8\,M_\odot$ behave similarly to classical bulges and early-type galaxies, while those with less massive black holes, particular the narrow-line Seyfert 1s, are consistent with pseudo bulges in late-type galaxies.  The host galaxy bulges, irrespective of whether they are classical or pseudo, follow the relatively tight inverse relation between effective radius and mean effective surface brightness of inactive classical bulges and ellipticals.  We argue that pseudo bulges experience recent or ongoing nuclear star formation.
\end{abstract}

\section{Introduction}
\label{sec:intro}

Supermassive black holes (BHs) are commonplace in the centers of massive galaxies, and their masses strongly correlate with the properties of the bulges of their host galaxies, including luminosity \citep{Kormendy1995}, stellar mass (\mbulge; \citealt{Magorrian1998}), and stellar velocity dispersion ($\sigma_\star$; \citealt{Ferrarese2000,Gebhardt2000}).  These scaling relations split into two branches \citep{Kormendy2013}: classical bulges and elliptical galaxies form a tight relation, while the relation for pseudo bulges exhibits markedly larger scatter and a different zero point.  The distinctly different trends observed for the two bulge types in terms of their BH-host correlations likely reflect their different evolutionary pathways \citep{Kormendy2004}.  The exact origin of these BH-host scaling relations is still a subject of lively research \citep{Menci2016, Thomas2019, Li2020, Terrazas2020}.  From an observational point of view, a number of pressing issues remain to be clarified.  Which came first, BH or galaxy?  Must the growth of the BH and its host be finely synchronized so as to preserve the small intrinsic scatter observed in the local scaling relations?  What is the cosmic evolution of the zero point and scatter of the scaling relations? Do they depend on the morphological type of the host or environment?  

An important step toward addressing these issues can be made by extending the BH-host scaling relations to active galaxies, wherein the BHs are {\it still}\ growing.  Toward this end, two ingredients are needed: BH masses and host galaxy parameters.  Fortunately, BH masses can now be estimated with reasonable ($\sim 0.3-0.4$ dex) accuracy  from single-epoch spectra for broad-line (type~1) active galactic nuclei (AGNs), particularly for $z \lesssim 0.75$ when the more reliable rest-frame optical lines can be accessible readily (e.g., \citealt{Vestergaard2006ApJ,Ho2015}).  By contrast, host galaxy parameters, especially for the bulge, are more challenging to obtain because of the bright glare of the AGN.  With considerable effort, stellar kinematics have been measured successfully for type~1 AGNs (e.g., \citealt{Greene2006, Bennert2015, Shen2015, Caglar2020}), but the samples are small, and $\sigma_\star$ becomes increasingly difficult, if not impossible, to measure for the most powerful quasars, especially at higher redshift.  The bulge stellar mass offers a promising alternative replacement of $\sigma_\star$ for investigating BH-host scaling relations.  At least insofar as classical bulges and ellipticals are concerned, \mbulge\ correlates just as tightly with \mbh\ as does $\sigma_\star$ \citep{Kormendy2013}.  Unlike stellar kinematics, however, \mbulge\ is more observationally tractable.  Securing a robust measurement of \mbulge\ for AGNs requires (1) high-resolution imaging with a stable point-spread function (PSF), (2) an effective analysis technique to deblend the host galaxy from the active nucleus and to decompose the bulge itself from other possible substructure in the host, and (3) sufficient color information to constrain the mass-to-light ratio ($M/L$).  Detailed two-dimensional (2D) image decomposition enables quantitative analysis of the structural components of quasar host galaxies (e.g., \citealt{Kim2008b,Veilleux2009ApJ,Schramm2013ApJ,Kim2017,Bentz2018ApJ,Li2020arXiv}).  

In this current work, we present Hubble Space Telescope (HST) observations of a small, carefully selected sample of nearby quasars designed to yield reliable photometry of the host galaxy, with the intent of estimating stellar masses. Special care is devoted to deriving accurate photometric and structural parameters for the bulge, for which it is of paramount importance to properly model strong bars, spiral arms, and other nonaxisymmetric structures.  Unlike earlier works \citep{Kim2008b,Kim2017}, here we provide rest-frame optical colors in order to mitigate possible variations in $M/L$, which can be considerable in light of the evidence for recent or ongoing star formation (e.g., \citealt{Husemann2017MNRAS,Kim2019, Zhao2019, Jarvis2020,Shangguan2020ApJ,Zhuang2020ApJ,Xie2021}).  We focus on low-redshift quasars in order to best characterize the local scaling relations for AGNs, laying the foundation for future investigations of possible cosmic evolution.  It is imperative to establish the true intrinsic scatter of the scaling relations, as well as to ascertain whether and how they might depend on the properties of the AGNs and the host galaxies.  For example, if substantial BH growth occurs during the AGN phase without synchronized star formation in the bulge, we expect the zero point or scatter of the scaling relations to depend on the mass accretion rate ($L_{\rm bol}/L_{\rm Edd}$) and/or color of the bulge. Or, if galaxy major mergers play a role in establishing the relations, their scatter might correlate with the presence of tidal features or local environment.  The dependence on bulge type and galaxy morphology has been well-established for inactive galaxies (e.g., \citealt{Kormendy2013,Greene2020}, and references therein), and a similar effort should be extended to their active counterparts.  While the sample considered here is far too small to definitively address all of these issues, it serves as a useful starting point.

Our HST observations and data reduction are described in Section~\ref{Sec:Data}.  We build a library of empirical PSFs performed bulge-disk decomposition in Section~\ref{Sec:GALFIT}.  Section~\ref{Sec:Res} analyses the main results on morphological classification, stellar mass estimation, and calculation of BH masses.   Implications are discussed in Section~\ref{Sec:Disc}, and the main conclusions are summarized in Section~\ref{Sec:Con}.  This work adopts the following parameters for a $\Lambda$CDM cosmology: $\Omega_m = 0.308$, $\Omega_\Lambda = 0.692$, and $H_0 = 67.8\,\mathrm{km\,s^{-1}\,Mpc^{-1}}$ \citep{Planck2016AA}.

\begin{figure}[htbp]
\begin{center}
\includegraphics[height=0.3\textheight]{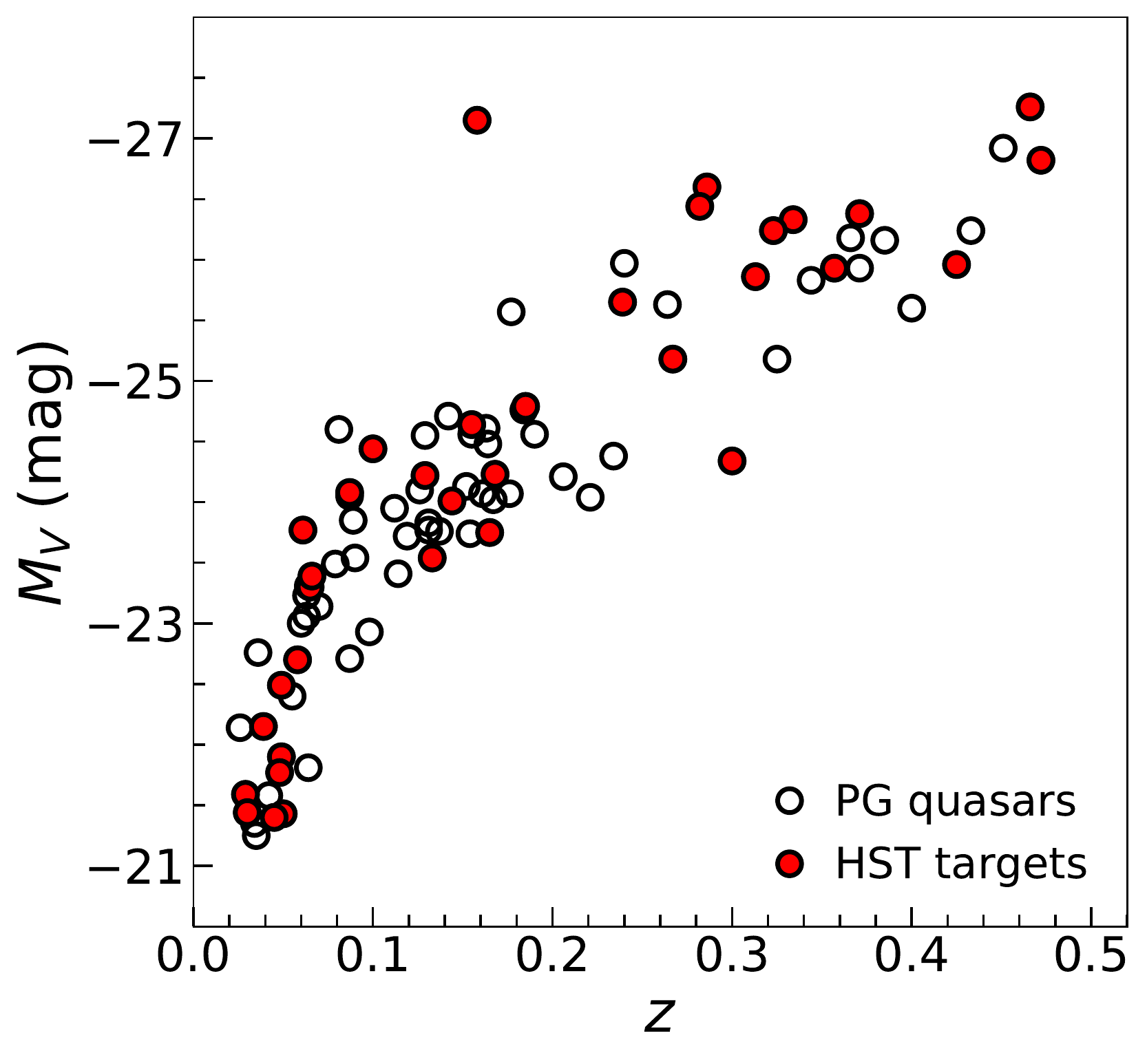}
\caption{The distribution of $V$-band absolute magnitude versus redshift for the 87 low-redshift ($z<0.5$) PG quasars.  The filled red circles are the 35 targets observed by our snapshot survey.}
\label{Fig:T1}
\end{center}
\end{figure}

\section{Data} 
\label{Sec:Data}

\subsection{Sample and Observations} 
\label{sec:smp}

Our parent sample consists of the 87 low-redshift ($z<0.5$) optically/UV-selected quasars from the Palomar-Green (PG) survey \citep{Schmidt1983,Boroson1992}.  Although the PG survey is incomplete due to large photometric error \citep{Goldschmidt1992}, this representative sample of nearby, bright quasars has been extensively studied for decades.  The availability of line widths for the broad H$\beta$ emission line from single-epoch spectroscopy \citep{Boroson1992,HoKim2009} and spectrophotometry for the optical continuum \citep{Neugebauer1987ApJS} enables BH masses to be estimated from conventional empirical calibrations (e.g., \citealt{Vestergaard2006ApJ}; \citealt{Ho2015}).  Robust bolometric luminosities, and hence Eddington ratios, can be derived from comprehensive broadband spectral energy distributions \citep{Ho2008}.  Figure~\ref{Fig:T1} shows the distribution of $V$-band absolute magnitude versus redshift for the sample.

We undertook a snapshot program (proposal 12903; PI: L. C. Ho) using the HST Wide Field Camera 3 (WFC3; \citealt{Dressel2019}) to obtain high-resolution rest-frame optical images for a significant subset of the PG quasars.  Snapshot programs cannot guarantee that any specific target will be observed, instead drawing from a pre-designated list of candidates to fill in gaps in the HST schedule to obtain relatively short observations of typical duration 45 minutes or less (including all overheads).  Our program achieved a success rate of $\sim 40\%$, securing useful data for 35 sources randomly selected from the original sample of 87 quasars.  The final sample forms a representative subset of the original sample (Figure~\ref{Fig:T1}).  Details of the observed targets and the observations are summarized in Table~\ref{tab:basic}.  

Each target was observed with two filters, which roughly correspond to rest-frame $B$ band and $I$ band, with the intent of providing a sufficiently wide color leverage to constrain the $M/L$ of the stellar continuum in order to estimate the stellar mass of the host galaxy.  Each pair of filters was customized carefully according to the redshift of the source to avoid, to the extent possible, contamination from strong emission lines.  In general each object was observed with both the UVIS and IR channels of WFC3, except for the nine sources with $z \lesssim 0.05$, for which both filters fall in the UVIS channel.  The UVIS detector has a pixel scale of $0\farcs0400 \times 0\farcs0398$ and a point-spread function (PSF) with a full-width at half maximum (FWHM) of $\sim 0\farcs08$.  The IR channel has a pixel scale of $0\farcs135 \times 0\farcs121$ and a PSF with ${\rm FWHM} \approx 0\farcs2$.  To reduce the readout overheads, we employ the {\tt UVIS2-M1K1C-SUB} subarray for the UVIS detector and the {\tt IRSUB512} subarray for the IR detector, which yield, respectively, a field-of-view of $40\arcsec \times 40\arcsec$ and $69\arcsec \times 62\arcsec$, adequate to cover the target and sufficient nearby field for background subtraction.  Exposure times varied.  Every target was observed with a long exposure with a total integration time of $\sim 200-500$~s in rest-frame $B$ and $\sim 150-300$~s in rest-frame $I$.  We use a three-point and four-point dither pattern for the UVIS and IR observations, respectively, to facilitate rejection of cosmic rays and bad pixels, as well as to better sample the PSF.  Given the brightness of the quasar at short wavelengths, saturation in the center is unavoidable in the long UVIS exposure.  We acquired an additional short ($\sim$ 6--100~s) UVIS exposure for the purposes of replacing the saturated nuclear region of the long exposure.  The IR images, taken in {\tt MULTIACCUM} mode with {\tt STEP25} timing sequence ($\mathrm{NSAMP}=6$), are not saturated.

Most of the observations were executed in the standard {\tt FINE LOCK} mode, but four targets (PG~0050+124, 1202+281, 1226+023, and 1244+026) were observed in {\tt GYRO} mode after a failure in guide star acquisition.  This resulted in significant degradation of the pointing precision during the observations, prompting us to apply corrective procedures described in Appendix \ref{Append:WCS}.

\subsection{Data Reduction}
\label{subsec:DR}

We reprocessed the calibrated, flat-fielded science files using the auxiliary data sets acquired from the Hubble Data Archive\footnote{\url{https://hla.stsci.edu/}} on 25 January 2016.  We used {\tt AstroDrizzle} \citep{Gonzaga2012}, largely with default parameters, to process the single-frame short exposure and to combine the dithered frames of the long exposure to remove cosmic rays and correct for geometric distortion.  The dithered sub-exposures are weighted by their maps of effective exposure time during the drizzle combination.  The pixel scale of the combined image is set to $0\farcs06$ for the IR channel and $0\farcs03$ for the UVIS channel to achieve Nyquist sampling.  The final images have a size of $54\arcsec \times 54\arcsec$ for the IR channel and $36\arcsec \times 36\arcsec$ for the UVIS channel.  The saturated nucleus of the long UVIS exposure was replaced by the corresponding pixels from the unsaturated short exposure, which was aligned to the former using {\tt AstroDrizzle}.  Depending on the level of saturation of the cores, we usually replace a circular region with radius $0\farcs3$.

\subsection{Background Estimation}
\label{sec:bkg}

Although {\tt AstroDrizzle} performs background removal, we take additional steps to verify the accuracy of the background determination and quantify its uncertainty.  While the pixel-to-pixel variation of the background emission is dominated by Poisson noise, often the main uncertainty of the background comes from larger scale fluctuations.  We first use the deeper $I$-band image to create a source mask from the segmentation image produced by {\tt SExtractor} \citep{Bertin1996}.  The $I$-band mask is then applied to the $B$-band image.  As in \cite{Li2011}, the mask is expanded according to the flux of the source.  Special care is paid to mask the region of the quasar and its host galaxy, whose total radial extent is determined from inspection of its one-dimensional (1D), azimuthally averaged surface brightness profile.  If the quasar is sufficiently strong, the mask for its diffraction spikes needs to be adjusted manually.  In order to identify the spatial scale and amplitude of the background variation, we randomly sample the unmasked pixels $10^4$ times with boxes of increasing size.  For each box size, we calculate the RMS of the sampled median background values. The scale of the background fluctuation is approximated by the box size above which the RMS converges, and the uncertainty of the background is given by the RMS value \citep{Li2011,Huang2013}.  A constant background, the median of the sampled values, is subtracted from the image.

\begin{figure*}[htbp]
\begin{center}
\includegraphics[width=0.9\textwidth]{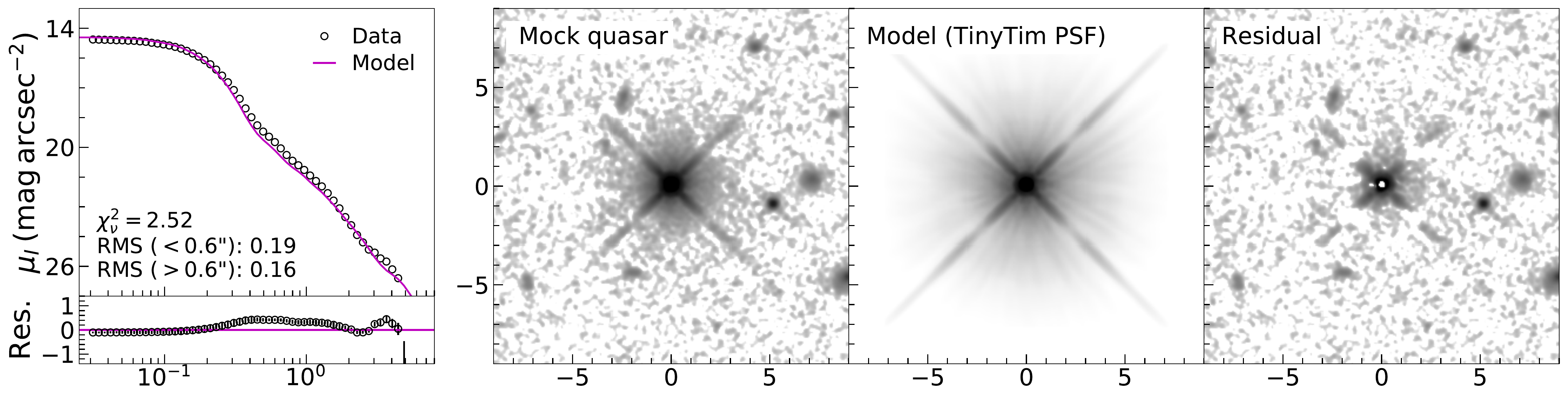}
\includegraphics[width=0.9\textwidth]{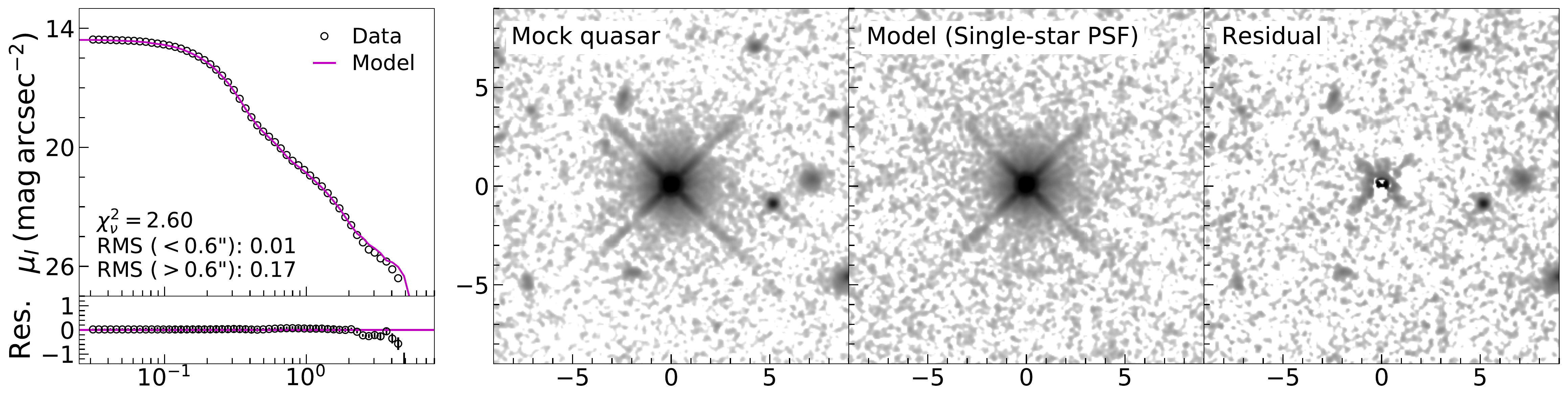}
\includegraphics[width=0.9\textwidth]{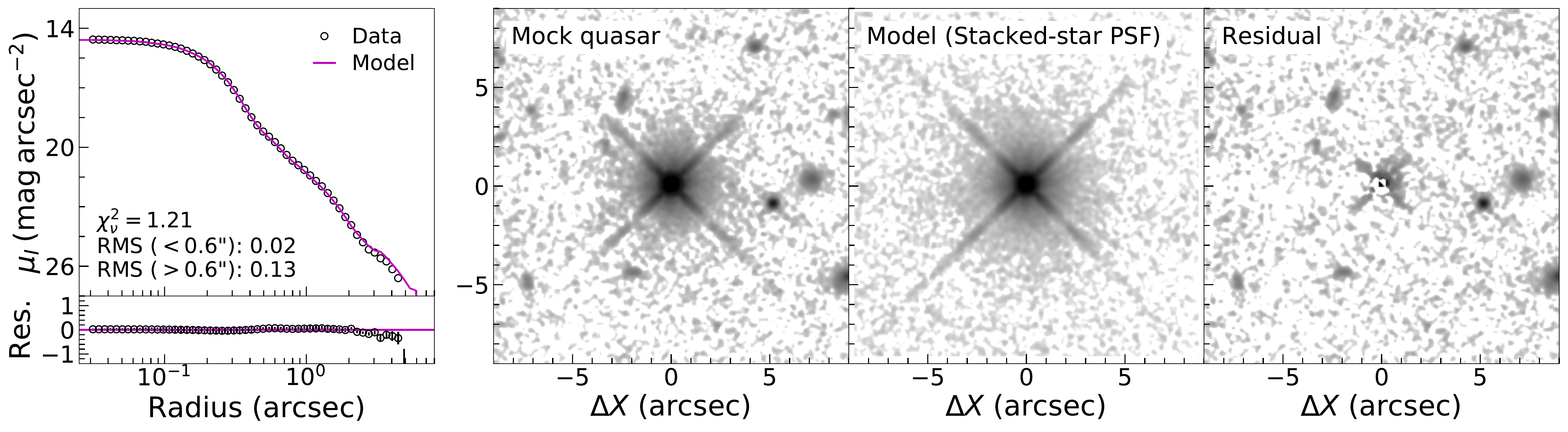}
\caption{Using a single star to mimic the WFC3 $I$-band observations of a quasar without a host galaxy, we fit the mock quasar with a PSF generated from (top) {\tt TinyTim}, (middle) a single star, and (bottom) stacking a large number of stars.  From left to right, the panels show the 1D profile, the observed image of the mock quasar, the best-fit \galfit\ model, and the residuals.  The 1D azimuthally averaged profile shows the original data (black circles) and the best-fit model (magenta solid line). The reduced $\chi^2$ from \galfit\ as well as the RMS of the 1D surface brightness residuals at radius below and above 0\farcs6 are shown in the lower-left corner.  All images are on an asinh stretch.
}
\label{Fig:PSF_dif}
\end{center}
\end{figure*}

\section{Image Decomposition}
\label{Sec:GALFIT}

The bright nucleus of type~1 AGNs poses a severe challenge for quantitative measurements of the stellar light of the underlying host galaxy, especially so for the bulge component, which is affected most acutely by contamination from the central point source \citep[e.g.,][]{Mclure2002,Peng2006b, Greene2008,Kim2008a,Kim2008b,Jiang2011,Kim2017}.  We use the 2D fitting code  \galfit\ \citep[version 3.0;][]{Peng2002,Peng2010} to simultaneously model the active nucleus and the host galaxy with parametric models. \galfit, a nonlinear fitting algorithm that uses Levenberg-Marquardt least-square minimization \citep{Press1992}, has been widely used to study the morphologies of galaxies, by decomposing their substructures using various PSF-convolved parametric functions to account for the bulge, disk, bar, and even spiral arms and tidal features \citep{Peng2010}.  A point-like nucleus can be added when an AGN is present.

We model the bulge, disk, and bar with the \sersic\ (1968) function 
 
\begin{equation} \label{eq:sersic}
\Sigma(r) = \Sigma_e\, \exp\left[-\kappa \left(\frac{r}{R_e}\right)^{1/n}-1\right],
\end{equation}

\noindent
where $\Sigma_e$ is the surface brightness at the effective radius $R_e$, $n$ is the \sersic\ index, which controls the concentration of the profile, and $\kappa$ is related to $n$ by the incomplete-gamma function \(\Gamma\left(2n\right)=2\gamma\left(2n,\kappa\right)\) \citep{Graham2005}.  When $n$ is large, the \sersic\ function has a steep inner profile and an extended outer wing; when $n$ is small, the \sersic\ function has a shallow inner profile and a steep truncation at large radius.  When fit as single-component entities, elliptical galaxies have $n \approx 3-8$ (e.g., \citealt{Huang2013}), classical bulges exhibit $n \approx 1.5-4$ (e.g., \citealt{Gao2018,Gao2020}), and pseudo bulges have $n \approx 1-3$ \citep{Gao2020}.  Disks are well-described by an exponential ($n \approx 1$; \citealt{Freeman1970}), at least over the radial extent not modified strongly by inner or outer breaks \citep[e.g.,][]{Pohlen2006,Li2011,Gao2017,Gao2019}, while bars usually can be fit with an $n=0.5$ profile \citep{Freeman1966,deJong1996b}.  \galfit\ employs coordinate rotation to produce realistic-looking spiral arms, and Fourier modes can mimic lopsidedness and a variety of non-axisymmetric tidal features \citep{Peng2010}.

\subsection{Point-spread Function}
\label{sec:psf}

Accurate knowledge of the PSF is critical to decomposing the bright nucleus of an active galaxy, a prerequisite to deriving reliable measurements of the host.  While the PSF of HST is much more stable than that of ground-based telescopes, it still varies with time due to telescope breathing and changes in instrument focus, and with position across the field due to optical distortion and the properties of the detector.  Ideally, an empirical PSF model can be obtained by observing a bright star with a spectrum similar to that of the science target, observed close in time to and using the same observing strategy as the science target \citep{Biretta2012,Mechtley2014}.  This option was not possible for our observations, in view of the severe scheduling limitations of the snapshot program.  Moreover, our decision to use subarrays to minimize readout time further significantly reduced  the field-of-view of the observations, to such an extent that hardly any of the images contains useful foreground stars that can serve as a PSF model.  In the absence of a dedicatedly observed PSF, many studies have utilized synthetic PSFs ({\tt TinyTim}; \citealt{Krist2011}) to analyze HST images of AGN host galaxies (e.g., \citealt{Greene2008,Kim2008a,Jiang2011,Kim2017}).  Here, we explore another strategy, by constructing an empirical PSF using stars extracted from archival images observed non-simultaneously in other HST programs, but otherwise employing the same detector-filter combination and dithering strategy.  We stack a number of stars to boost the signal-to-noise ratio.  For each detector-filter combination used in our program, we searched the HST archive for useful images that contain suitably isolated, unsaturated stars that are at least one-third as bright as the corresponding quasar. We do not include stars with close contamination (extended sources or bright point sources) within $\sim 3\arcsec$, where the diffraction spike is prominent. We use {\tt AstroDrizzle} to process the images of the stars with exactly the same procedure and pixel scale as the science images. Each image, typically $\sim 30\arcsec$ on a side, is inspected carefully for possible faint contaminating sources, which, when present, are replaced manually by random Gaussian noise using the standard deviation measured from the nearby background.  The final PSF for each filter, a median-stacked combination of typically $\sim 10-65$ stars, is only slightly ($\lesssim 10\%$) broader than the profile of individual stars.

To test the effectiveness of our strategy, we choose a single, bright star having roughly similar brightness as a typical quasar in our sample to serve as the mock image of a ``pure'' quasar nucleus (one lacking any underlying host galaxy), and we fit it with different PSF models to evaluate how well each can match the point source.  Figure~\ref{Fig:PSF_dif} compares the performance of (1) a standard synthetic PSF generated from {\tt TinyTim}, (2) an empirical PSF derived from a single star in the same field-of-view as the mock quasar with similar flux, and (3) an empirical PSF produced from stacking a large number of stars.\footnote{The star used to represent the mock quasar is not included in the stack.} The image of the mock quasar and the PSF models are convolved with a Gaussian kernel to achieve Nyquist sampling, as discussed in Section~\ref{sec:fit}.  Using \galfit\ to fit the mock quasar, we find that the stacked-star PSF performs slightly better than the single-star PSF, and both empirical PSFs are always superior to the {\tt TinyTim} PSF, as judged by the reduced $\chi^2$ reported by \galfit\ as well as the RMS of the residuals of the 1D profile. This conclusion holds for tests with different sets of stars on images of different bands (e.g., F438W, F814W, F105W, and F110W). We also explored the influence of the number of stars used in the stacking, finding, as a rule of thumb, that combining $\gtrsim 10$ stars produces the most optimal empirical PSF.

\subsection{Fitting Procedure}
\label{sec:fit}

PSF mismatch in the central pixels of bright nuclei presents a critical challenge for the analysis of the host galaxies of AGNs, especially those bright enough to be considered quasars.  The problem is exacerbated when the PSF is undersampled, as is the case for HST.  While employing dithering strategies can alleviate the problem, in practice this still proved insufficient for our purposes, especially in the absence of dedicated PSF observations (Section~\ref{sec:psf}).  Under these circumstances, one can mitigate the PSF mismatch by artificially broadening the PSF image and science image by a modest amount, by convolving them with a Gaussian kernel so that the PSF is Nyquist-sampled by the pixel scale of the detector \citep{Kim2008a}.  We adopt the same strategy for this study.  \galfit\ requires a ``$\sigma$ image'' to properly calculate $\chi^2$.  Following \cite{Casertano2000AJ}, the $\sigma$ image takes into account noise correlation due to the drizzling process of combining images and the different exposure times if the core of the nucleus is replaced.  We nevertheless confirm that the decomposition results, after accounting for systematic uncertainties, are not significantly different if we adopt the ``$\sigma$ image'' generated automatically by \galfit.

We first focus on the $I$-band data, which have higher signal-to-noise ratio than the $B$-band data.  The $I$ band also offers higher contrast between the host galaxy and the AGN.  Once a final, viable model is constructed for the $I$ band, we adopt the best-fit parameters and apply them to the $B$ band, adjusting only the flux amplitude of each component.  This approach, similar to that of \cite{Li2020arXiv}, ensures a consistent bulge-disk decomposition in both bands and mitigates severe PSF mismatch in the $B$ band, even if the $B$-band model may not be optimized (see Section~\ref{sec:unc}).  We use the same source mask to estimate the background.  We start with the simplest possible model---a point-like nucleus plus a single-component host galaxy---and slowly include more components (e.g., a disk or a bar) to account for additional substructure of the host galaxy if the data require.  The centers of all of the components are tied together.  Close companions, when sufficiently nearby, need to be fit simultaneously.  Whenever possible, we allow the \sersic\ index of the spheroidal component (elliptical galaxies and the bulges of disk galaxies) to be fit freely. However, as the  solutions are not always meaningful because of parameter degeneracies or strong systematic effects arising from PSF mismatch, we often need to fix the \sersic\ index to discrete values between $n = 1$ to $n = 4$ and choose that which produces the lowest residuals, with the range of acceptable solutions serving to bracket the uncertainties. Given the many systematic uncertainties in model selection, it is difficult to rely on formal statistical criteria (e.g., reduced $\chi^2$) to determine the goodness-of-fit.  Instead, we utilize the 1D and 2D residuals to inform us of the relative merits of different models and to decide on whether and in which manner more complex models need to be considered.  For example, spiral structure betrays the presence of a disk, and bisymmetric residuals often indicate that a bar is present.  We invoke coordinate rotation to capture obvious spiral arms, even if they do not significantly affect the photometric parameters of the bulges of inactive galaxies \citep{Gao2017}.  Bending modes can mimic tidal features effectively, and Fourier modes of order up to 3 can accommodate a variety of non-asymmetric structures in the outermost extended regions \citep[e.g.,][]{Kim2008b}.  In particular, the absolute amplitude of the $m=1$ Fourier mode ($A_L$) can effectively gauge the global asymmetry \citep{Peng2010}.  Quasar hosts with $A_L>0.2$ tend to be highly disturbed systems that are ongoing or late-stage mergers \citep{Kim2017}.

\begin{figure*}[htbp]
\centering
\includegraphics[width=0.9\textwidth]{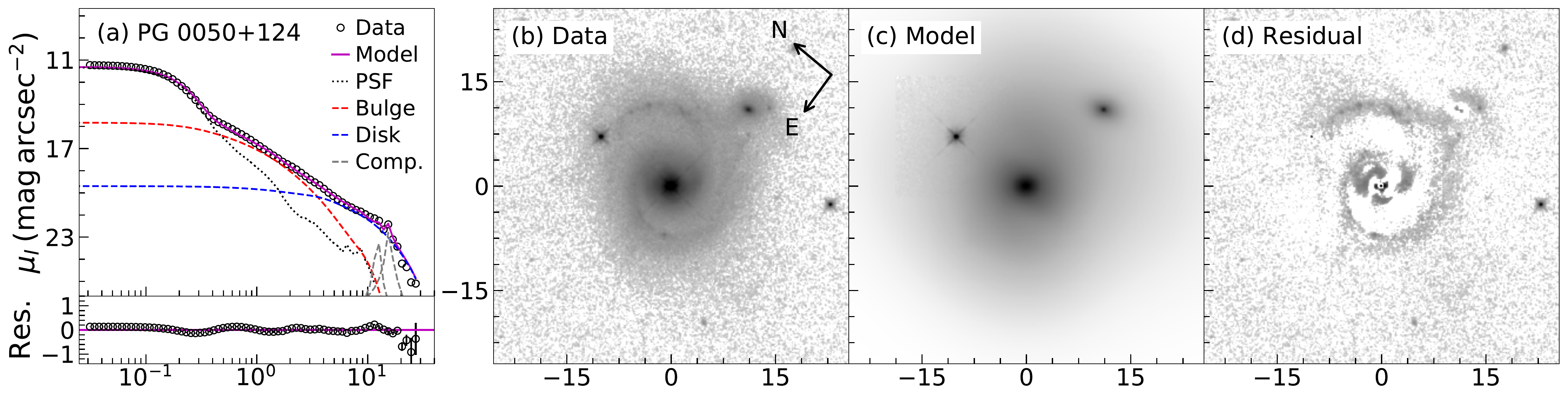}
\includegraphics[width=0.9\textwidth]{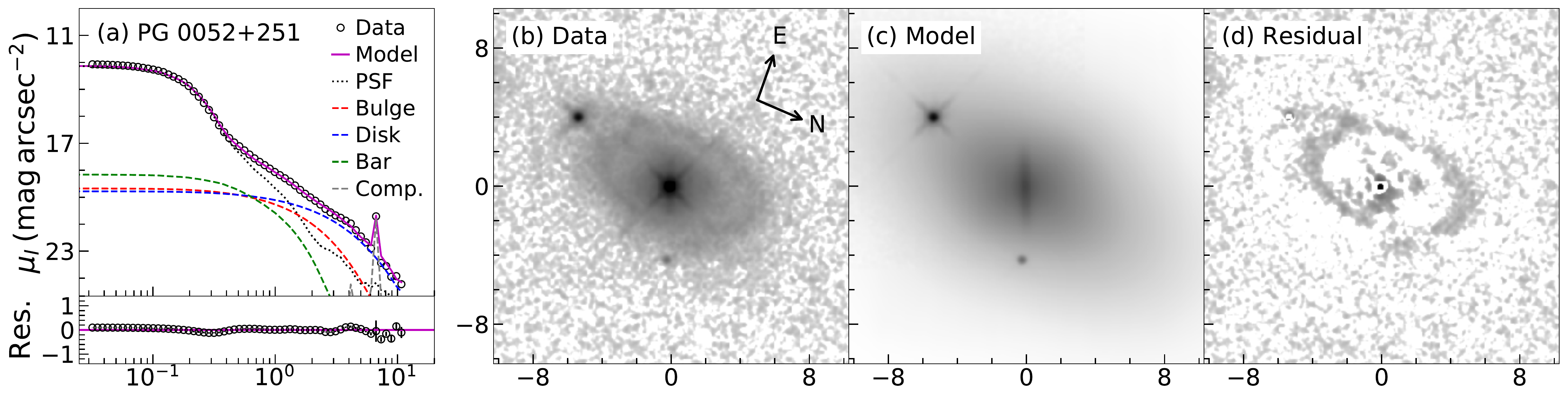}
\includegraphics[width=0.9\textwidth]{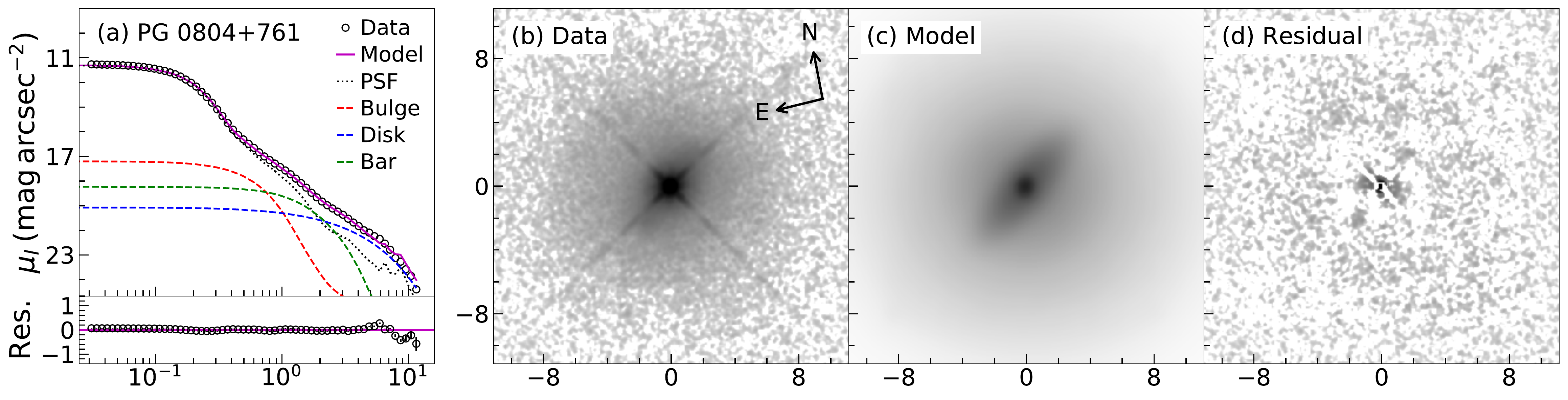}
\includegraphics[width=0.9\textwidth]{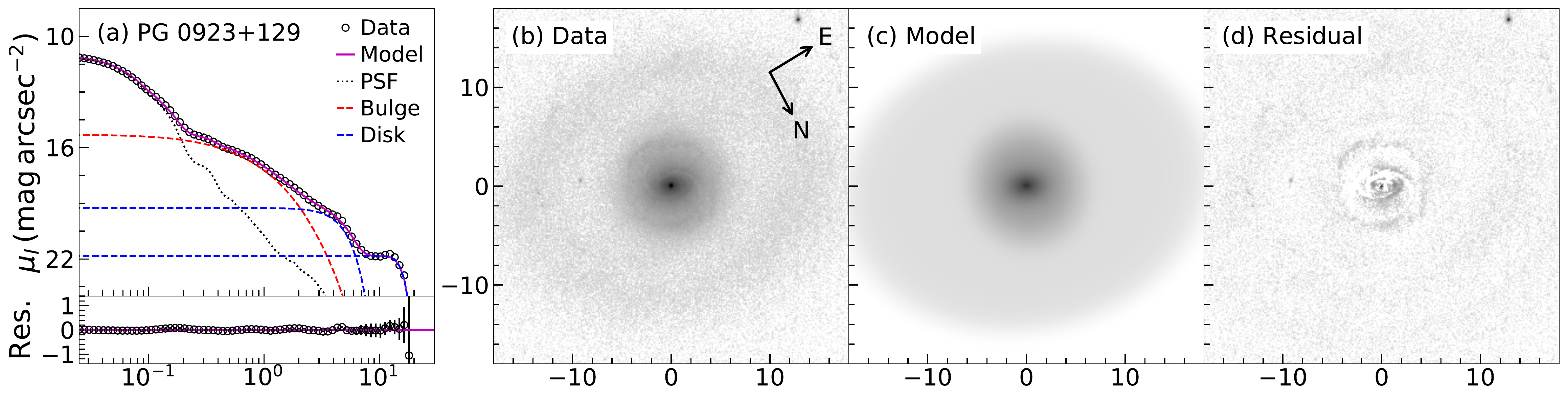}
\includegraphics[width=0.9\textwidth]{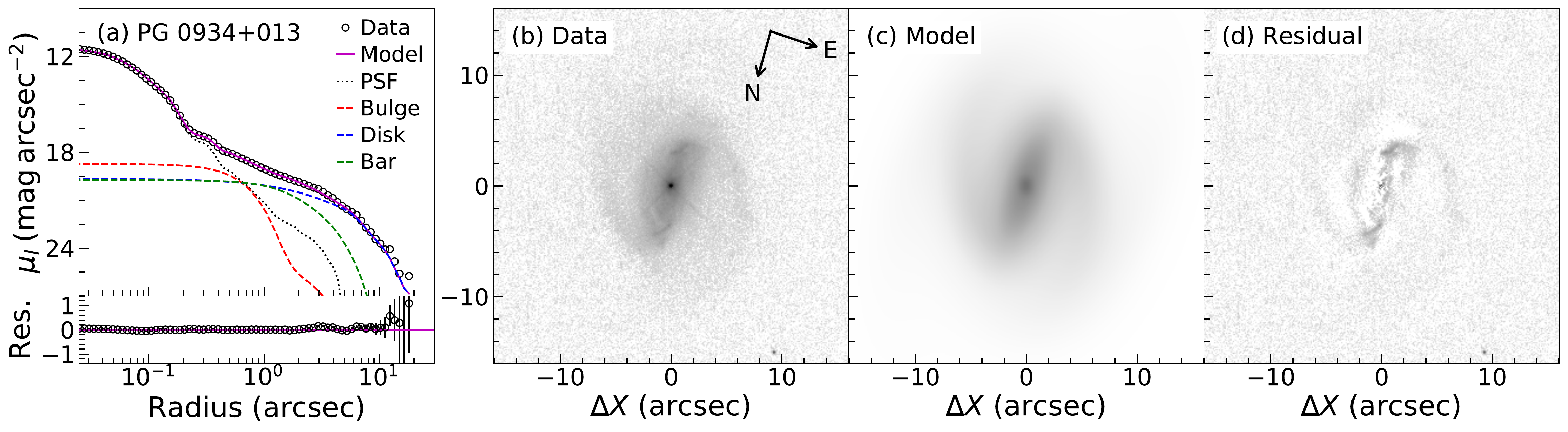}
\caption{ \galfit\ decomposition of $I$-band images, showing (a) the 1D profile, (b) the observed image, (c) the best-fit 2D model of the host (AGN component excluded), and (d) the residuals. The 1D azimuthally averaged profile shows the original data (black circles), the best-fit model (magenta solid line), and the subcomponents (PSF: black dotted line; bulge: red dashed line; disk: blue dashed line; bar: green dashed line; companion: gray dashed line). All images are on an asinh stretch.  The complete figure set (35 images) is available in the online journal.}
\label{Fig:gfI}
\end{figure*}

\begin{figure*}[htbp]
\centering
\includegraphics[width=0.9\textwidth]{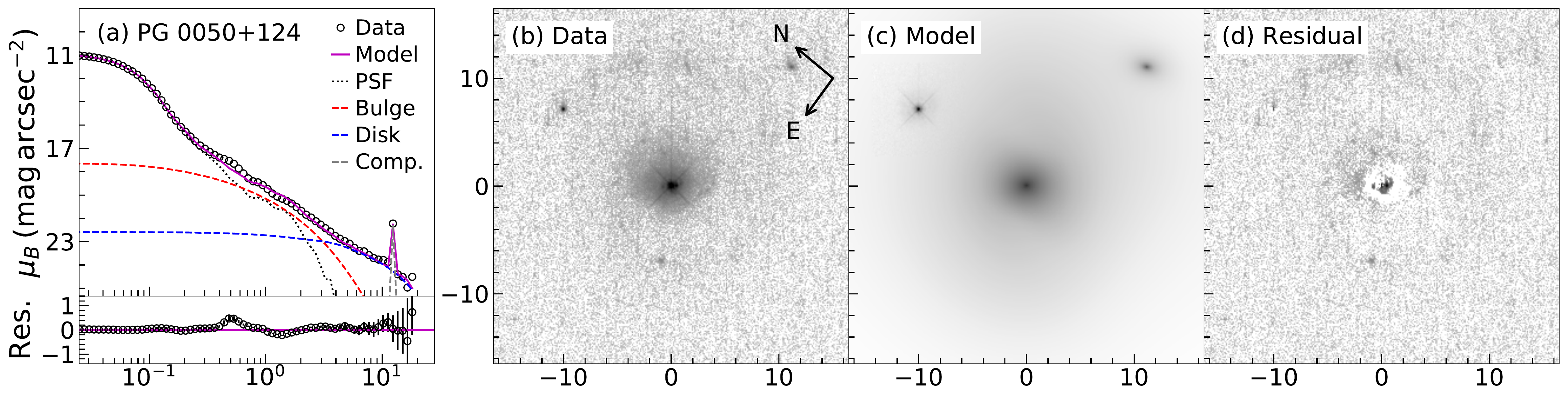}
\includegraphics[width=0.9\textwidth]{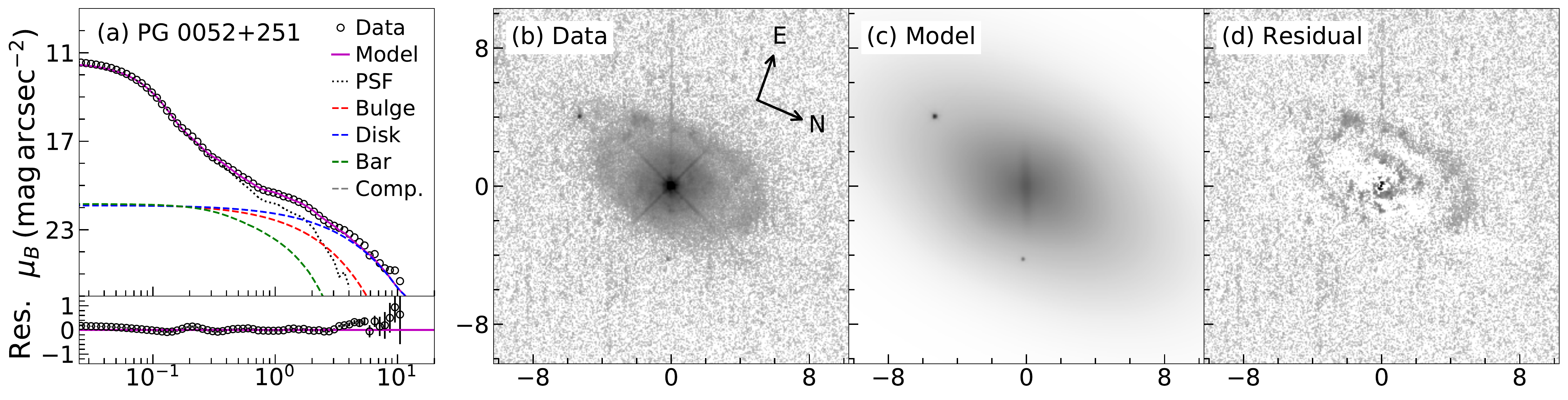}
\includegraphics[width=0.9\textwidth]{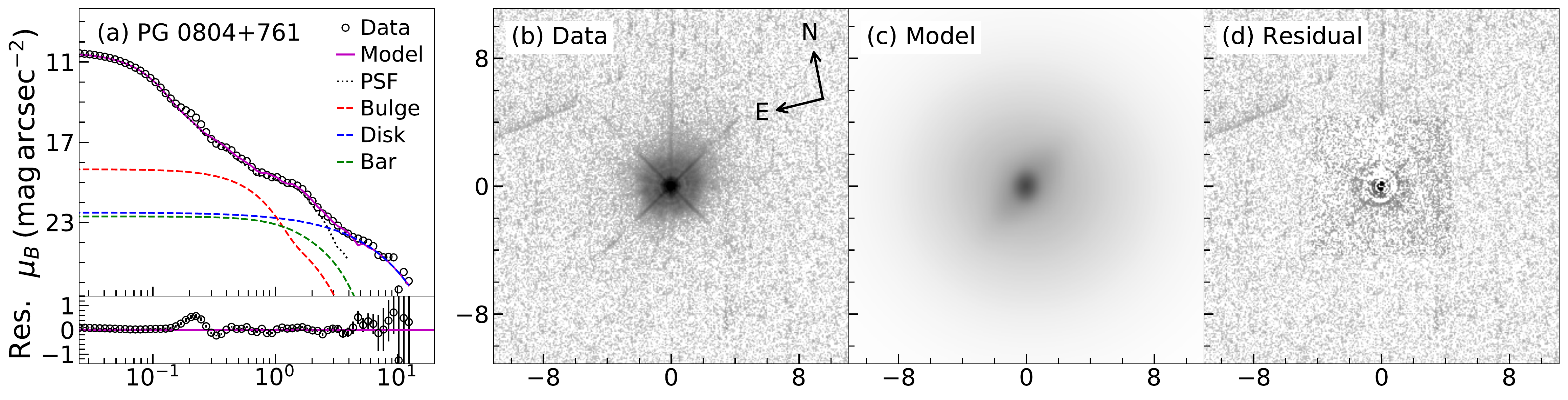}
\includegraphics[width=0.9\textwidth]{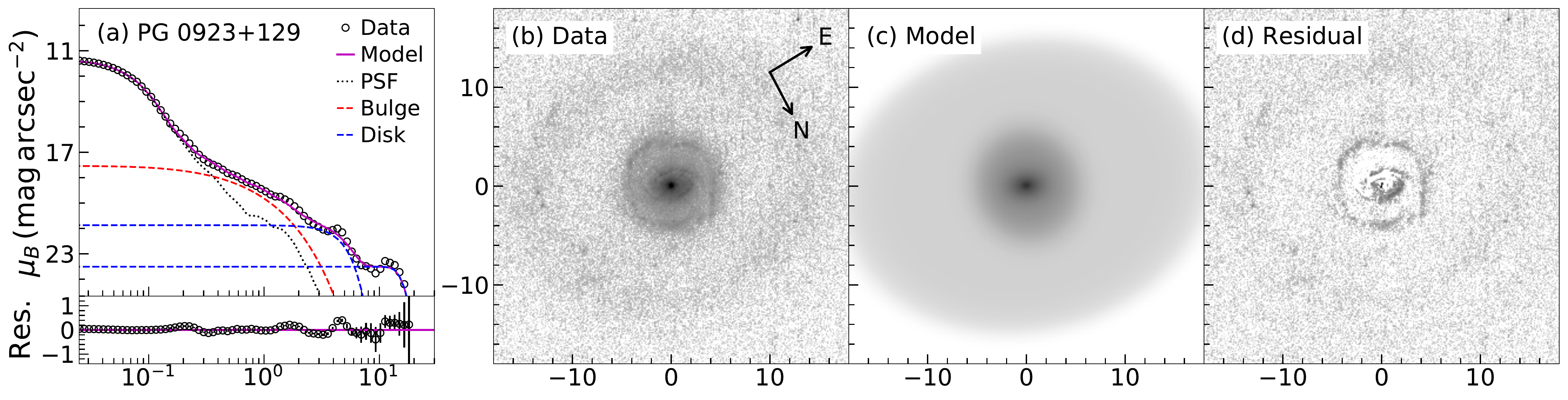}
\includegraphics[width=0.9\textwidth]{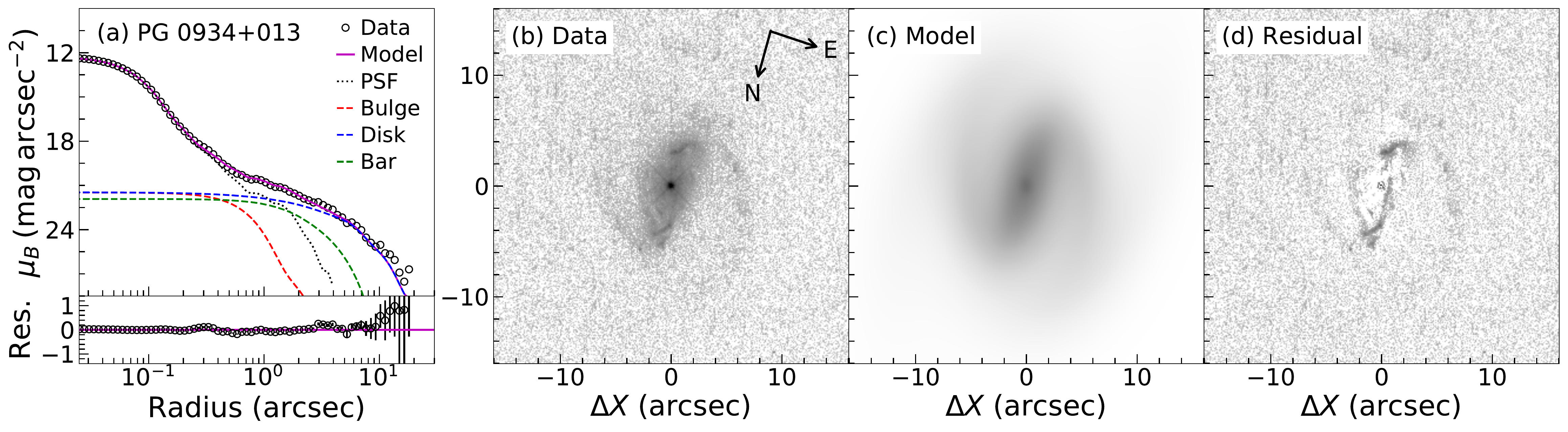}
\caption{ \galfit\ decomposition of $B$-band images, showing (a) the 1D profile, (b) the observed image, (c) the best-fit 2D model of the host (AGN component excluded), and (d) the residuals. The 1D azimuthally averaged profile shows the original data (black circles), the best-fit model (magenta solid line), and the subcomponents (PSF: black dotted line; bulge: red dashed line; disk: blue dashed line; bar: green dashed line; companion: gray dashed line). All images are on an asinh stretch.  The complete figure set (35 images) is available in the online journal.}
\label{Fig:gfB}
\end{figure*}

Figures~\ref{Fig:gfI} and \ref{Fig:gfB} show the fits of the $I$-band and $B$-band images, respectively, and Appendix~B provides notes on the individual objects.  The best-fit parameters are reported in Table~\ref{tab:galfit}.  

\subsection{Uncertainties}
\label{sec:unc}

The formal uncertainties reported by \galfit\ are usually small.  In practice, the error budget for most of the derived parameters are dominated by several sources of systematic uncertainty.  First, as described in Section~3.2, many of the fitting results, especially those that pertain to the bulge, are extremely sensitive to the PSF.  While we have made every effort to secure a reliable empirical PSF based on stacking a number of observed stars (Section~3.1), of course, the final PSF depends on the exact number and choice of stars.  To account for this factor, we repeat the fits with different realizations of the empirical PSF constructed from various combinations of observed stars.  Next, due to the limited field-of-view of our images, we opt to fix the background to its predetermined value (Section~2.3) in order to facilitate the convergence of the fit \citep{Kim2008a}, but we explore the impact of large-scale fluctuations of the background by repeating the fits with the background value perturbed around its RMS value.  Finally, as documented in Appendix~B, a major source of uncertainty occasionally arises from the ambiguity of model selection.  Because of the complexity of AGN host galaxies, the robustness of the fits must be scrutinized individually for each source, and often the ``best fit'' is based less on any rigorous statistical criteria than on a reasonable choice bracketed by a range of plausible choices.  The adopted range of acceptable solutions then defines the true uncertainty range.  The final uncertainties reported in Table~\ref{tab:galfit} reflect the quadrature sum of the statistical uncertainties from \galfit\ and the systematic uncertainties due to the (1) the PSF, (2) background subtraction, and (3) model selection.  

\begin{figure*}[htbp]
\centering
\includegraphics[height=0.3\textheight]{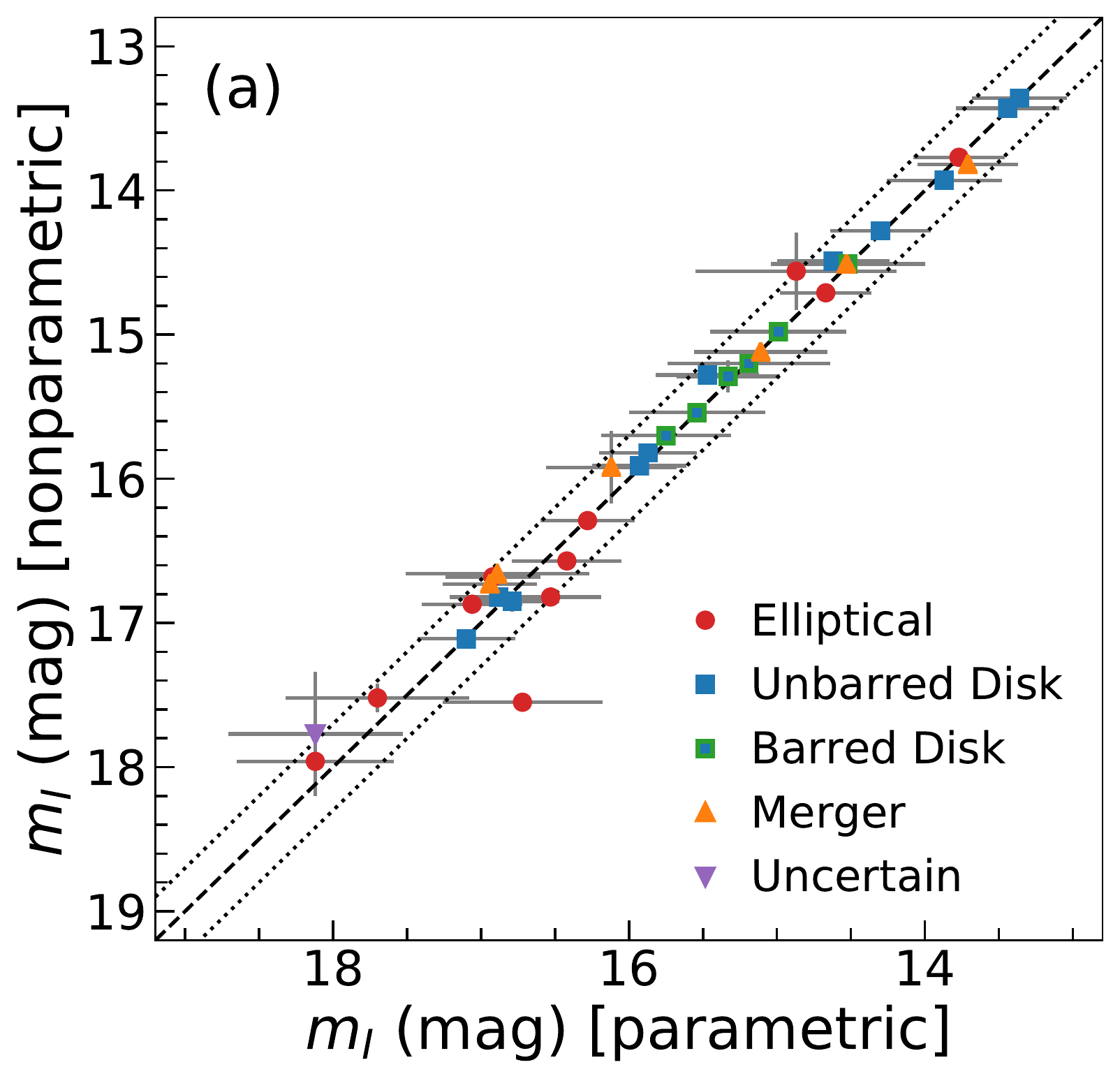}
\includegraphics[height=0.3\textheight]{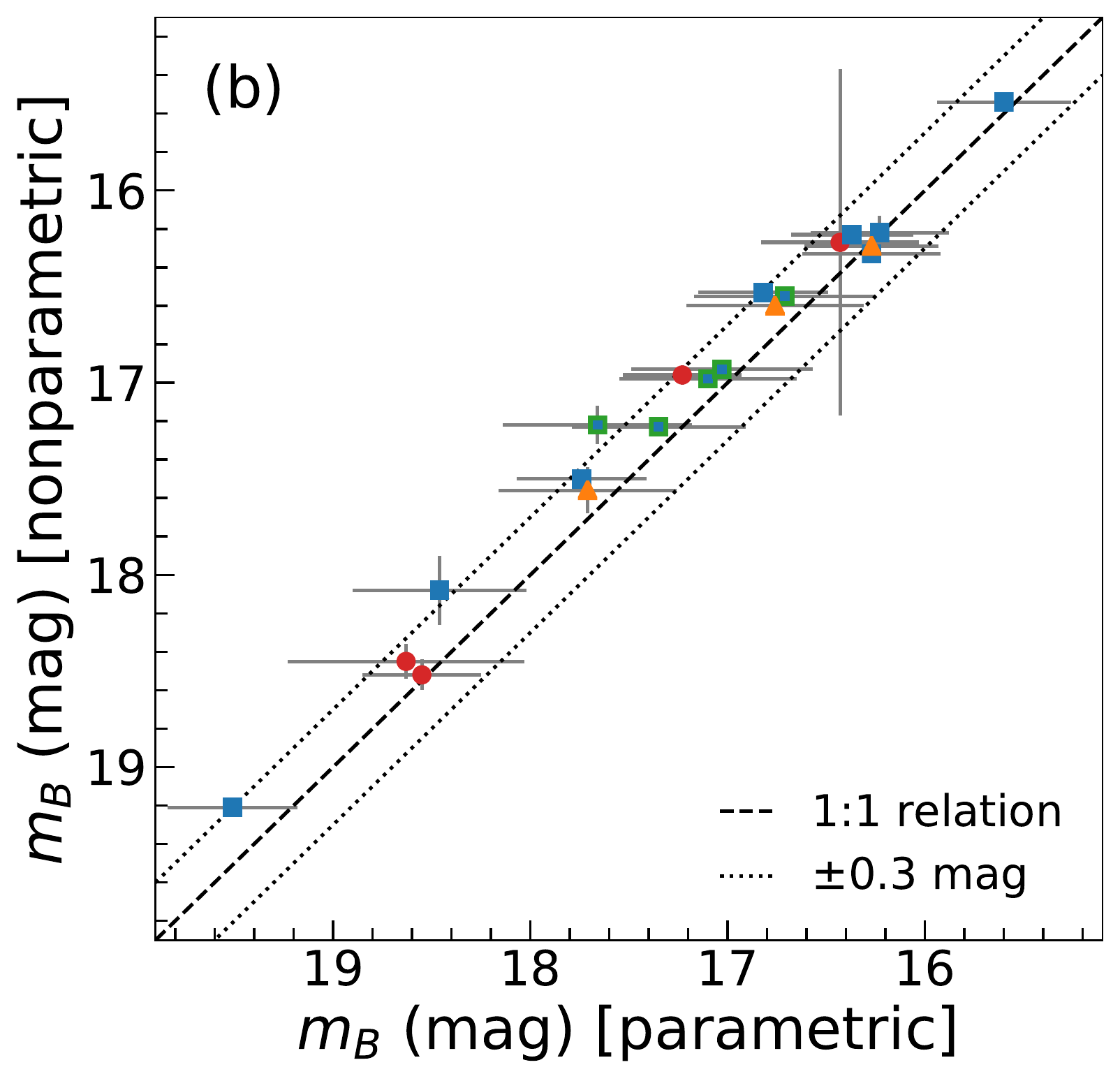}
\caption{Comparison of the integrated magnitude of the host galaxy based on parametric and nonparametric measurement for (a) $I$ band and (b) $B$ band.  Different morphological types are given in the legend.  The dashed line is the one-to-one relation, with the dotted lines indicating a difference of $\pm 0.3$ mag.  The $I$-band magnitudes show no systematic difference between the two methods, while in the $B$ band the parametric method yields median brightnesses that are 0.16 mag fainter than those from the nonparametric method (see text).
} 
\label{Fig:mpar}
\end{figure*}

Following \cite{Kim2008a}, we cross-check the decomposition results by comparing the parametrically derived total magnitudes with those measured nonparametrically (Figure~\ref{Fig:mpar}).  After removing the nuclear emission and close contaminating sources based on the overall \galfit\ model, we sum up all the remaining flux to estimate the total magnitude of the host galaxy.  The uncertainty of the nonparametric magnitude is estimated by adopting different PSF models and background subtraction.  The nonparametric magnitudes usually agree well with the sum of the parametric components, with the median difference between the parametric and nonparametric magnitudes and their standard deviation being $\Delta m = 0.01\pm0.19$~mag in the $I$ band.  PSF mismatch poses a more serious problem in the $B$ band and strongly affects the nonparametric magnitudes if the host galaxy is faint. For the 20 sources for which a meaningful comparison can be made, the $B$-band parametric magnitudes are fainter than the nonparametric magnitudes by $\Delta m = 0.16\pm0.13$~mag.  The parametric measurements, constrained by the structural parameters based on the $I$-band decomposition, may miss more extended emission in the $B$ band. Nevertheless, the difference is much smaller than the uncertainty of the integrated magnitudes, and in terms of stellar mass it amounts to only 0.07 dex, which is much smaller than the mass uncertainty.

\section{Results}  
\label{Sec:Res}

\subsection{Morphological Classification}
\label{sec:morph}

On account of the strong contamination by the AGN core, coupled with the large distances of the sources and the relative shallowness of the HST images, the morphology of the host galaxies is challenging to ascertain.  We base our morphological classification in part on our quantitative decomposition and in part on visual inspection of the original images and the residual images after subtracting the best-fit parametric model (Figures~\ref{Fig:gfI} and ~\ref{Fig:gfB}).  Except for a single case (Appendix~B), the 11 galaxies (31\% of the sample) that can be fit adequately with a single \sersic\ component are classified as ellipticals, even if in deep observations nearby ellipticals often exhibit more than one major photometric component \citep{Huang2013, Zhu2021}.  Although the visual classification is usually challenging when a source is at $z\gtrsim 0.2$, particularly when the nuclear emission overwhelms the starlight, the host galaxies, robustly detected, are sufficiently luminous and massive to be regarded securely as ellipticals.  Disk galaxies are recognizable as two-component systems with a \sersic\ bulge plus an exponential disk, which is often accompanied by an additional non-axisymmetric feature that can be readily identified with a bar.  Spiral features in the residual maps offer further supporting evidence for the presence of a disk.  Nearly half of the sample (49\%, 17/35) resides in disks, among them $\sim 35\%$ (6/17) that are barred.  Six objects (17\%) exhibit irregular substructure indicative of unambiguous ongoing or recent merger activity.  Throughout the paper, we mainly focus on galaxy major mergers with stellar mass ratio typically $\lesssim 0.3$.  PG~1202+281 has a close projected neighbor, but in the absence of detectable tidal features or redshift information for the nearby source, we do not consider it a merger system. The single-component hosts are systematically more luminous (median $M_I = -23.71$~mag) than those decomposed into two (bulge and disk) or three (bulge, bar, and disk) components (median $M_I = -22.66$~mag), consistent with the known luminosity difference between nearby ellipticals and spirals (e.g., \citealt{Burstein2005}).  This gives us some confidence that the ellipticals have been properly classified.  The mergers are the most luminous, with median $M_I = -24.21$~mag, not too dissimilar from the ellipticals.

\begin{figure*}[htbp]
\centering
\includegraphics[height=0.3\textheight]{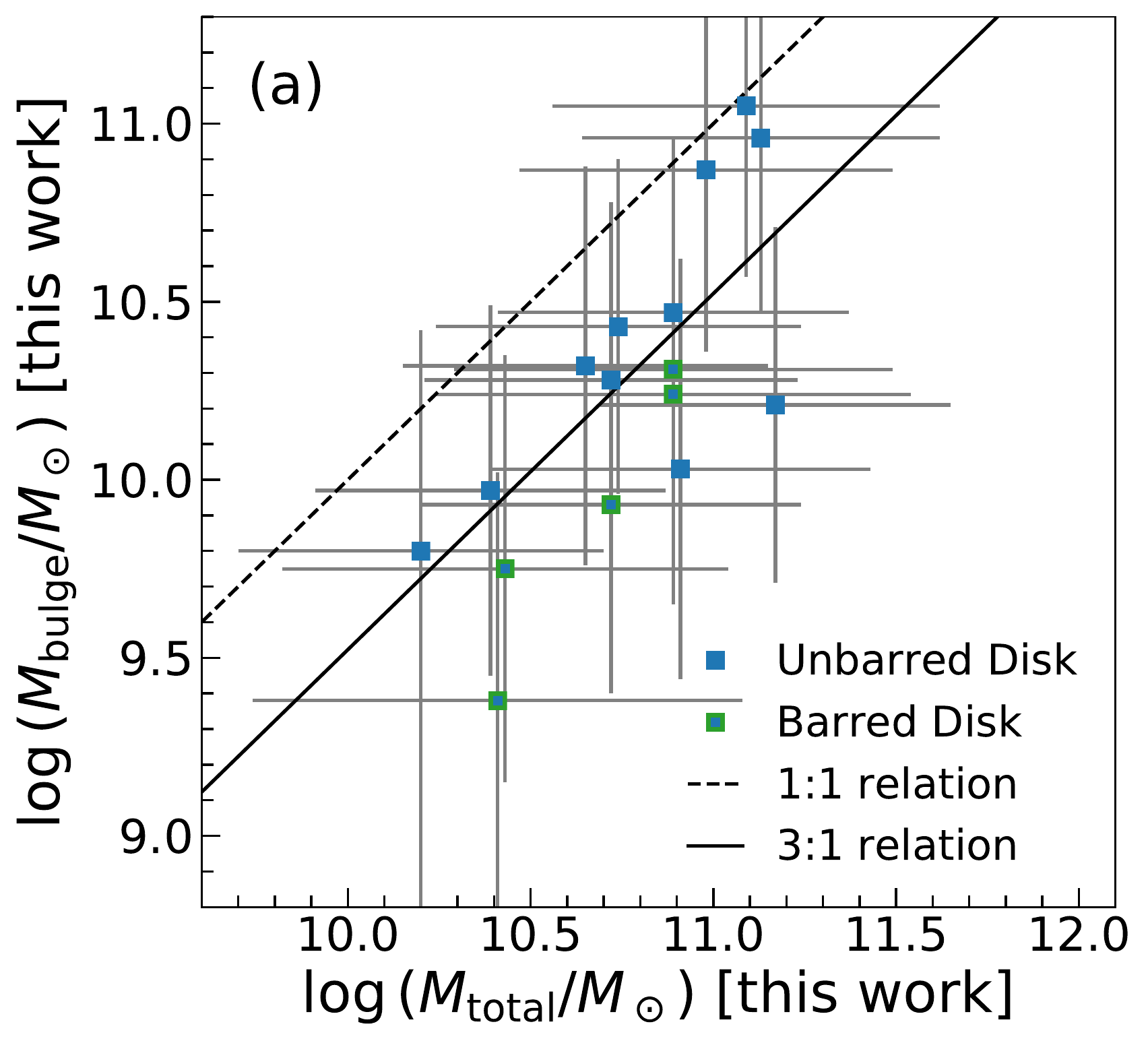}
\includegraphics[height=0.3\textheight]{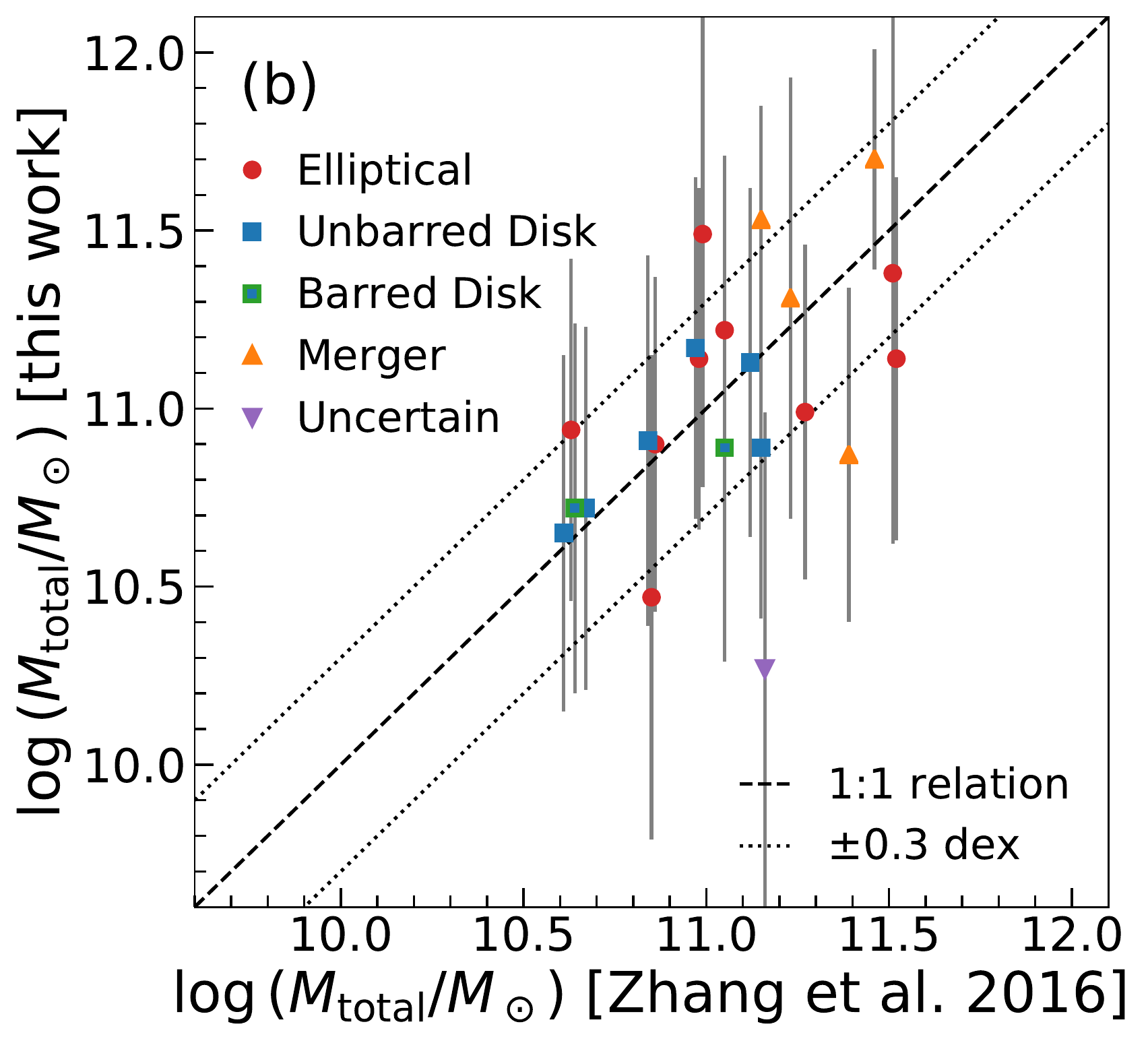}
\caption{Comparison of the total stellar mass of the host galaxies derived in this work with (a) the stellar mass of the bulges of the disk galaxies and (b) the total stellar mass from \cite{Zhang2016ApJ}.  The dashed line indicates the 1:1 relation.  In panel (a), the solid line indicates the 3:1 relation; the bulge mass is always smaller than the total stellar mass of the system.  In panel (b), the dotted lines mark $\pm 0.3$~dex; the two sets of estimates are mostly consistent within the 0.3~dex uncertainty.} 
\label{Fig:cmpb}
\end{figure*}

\subsection{Stellar Mass Estimation}
\label{sec:mstar}

We estimate the total stellar mass of the host galaxy, and separately for the bulge alone if present, from its rest-frame, extinction-corrected $I$-band magnitude and $B-I$ color, using the color-dependent $M/L$ prescribed by \cite{Bell2001}.  To convert the measured HST-based magnitudes to rest-frame magnitudes in the Johnson $B$ and $I$ bands, we adopt \cite{Bruzual2003} stellar population synthesis models covering a grid of stellar ages provided by \hst\ calibration reference data.\footnote{\url{http://etc.stsci.edu/etcstatic/users_guide/1_ref_2_spectral_distribution.html}}  The models assume an exponentially decaying star formation history with an e-folding time of 0.6~Gyr, solar metallicity, and a \cite{Chabrier2003} stellar initial mass function (IMF).  The model spectra are redshifted to the observed frame of each target, corrected for foreground extinction using the extinction values of \cite{SF2011} and the Galactic extinction curve of \cite{Cardelli1989}, and then convolved with the respective HST filter response function using \texttt{synphot}\footnote{\url{https://synphot.readthedocs.io}}.  The model spectrum that provides the closest match to the observed color and its associated uncertainty is used to calculate the K-correction.  Typical ages for the bulge component range between 1 and 12 Gyr, overlapping with but systematiacally larger than the ages of 1--5 Gyr that characterize the entire galaxy.  The uncertainty of the K-correction is typically small ($\lesssim 0.05$ mag; \citealt{Kim2017}).  The derived quantities are not sensitive to the star formation history of the adopted stellar synthesis model.  Whether we use the simple stellar population model or the exponential decaying model with an e-folding time 1.0 Gyr, our derived bulge and host galaxy masses varies by typically $\lesssim 0.1$ dex.

The calculations of \cite{Bell2001} are based on the IMF of \cite{Salpeter1955}.  Scaling the stellar masses by a factor of 0.61 \citep{Madau2014} to match the IMF of \cite{Chabrier2003},
 
\begin{equation}\label{eq:mass}
\mathrm{\log}\left(\frac{M_*}{M_\odot}\right) = -0.4 \, (M_I - M_{I, \odot}) + 0.439 \, (B-I) - 0.609,
\end{equation}
 
\noindent
where $M_{I, \odot}=4.07$ mag is the $I$-band absolute magnitude of the Sun \citep{Blanton2007}.  The final uncertainties of the stellar masses are significant (median $\sim 0.5$ dex; Table~\ref{Tab:tab_derived}), accounting not only for the $\sim$0.3~dex uncertainty of $M/L$ \citep{Conroy2013ARAA}, but also for the significant uncertainty associated with the photometric decomposition.  Our subsequent analysis will use the results based on the parametric magnitudes, except for the hosts of objects classified as mergers, for which the nonparametric magnitudes should be more reliable when the nuclei are properly removed.  The concept of a ``bulge,'' of course, is not meaningful in an ongoing merger.  As all the mergers in our sample appear to be quite substantial, for simplicity we assume that the final product of the merger will be an elliptical, or at the very least a highly bulge-dominated galaxy, even if a disk can reform (e.g., \citealt{Peschken2020MNRAS}).  

Despite the large uncertainties of the stellar masses, it is reassuring that the bulge masses are always smaller than the total masses of the host (Figure~\ref{Fig:cmpb}a).  The median bulge-to-total stellar mass ratio is $\sim 0.4$.  While this may seem an obvious point, it does offer some measure of reassurance that our masses are not unreasonable despite the challenges associated with estimating $M/L$ for the bulge and the galaxy separately, which can occasionally produce bulge masses that formally exceed those of the whole galaxy \citep{Bentz2018ApJ,Li2020arXiv}.  Prior to this study, the work of \cite{Zhang2016ApJ} has provided the largest collection of (total) stellar masses for the PG quasars, derived from assembling heterogeneous, literature-based HST photometry of 55 sources assuming a fixed $M/L$ for the optical or near-IR.  We compare our host galaxy stellar masses with those of \cite{Zhang2016ApJ} in Figure~\ref{Fig:cmpb}b, for the subset of 22 overlapping sources.\footnote{If two stellar masses are given in Table~1 of \cite{Zhang2016ApJ}, we use their average value.  We scale their masses to our adopted IMF \citep{Chabrier2003} and cosmology \citep{Planck2016AA}.} The two sets of stellar masses show overall agreement, with most sources lying within an envelope of $\pm 0.3$ dex.  The most deviant points are PSF-dominated sources and mergers.  The most striking example is PG~0953+414, for which our stellar mass of $M_* = 1.9 \times 10^{10}\,M_\odot$ is $\sim 10$ times lower than that reported by \cite{Zhang2016ApJ}.  The optical photometry used by \cite{Zhang2016ApJ} originated from the analysis of an HST WFPC2/F675W image by \cite{Hamilton2002ApJ}, who seem to have severely overestimated the flux of the host \citep{Kim2017}, likely on account of the extremely dominant nuclear emission of PG~0953+414.  Similarly, our stellar mass for PG~1700+518 is a factor of $\sim 3$ lower than that reported by \cite{Zhang2016ApJ}, who based it on an HST NICMOS/F160W observation.  \cite{Bentz2018ApJ} analyzed the same image for this source and found a stellar mass similar to ours.

\subsection{Black Hole Mass Estimation}
\label{sec:mbh}

Following \cite{Ho2015}, the BH mass is derived as

\begin{equation}
\mathrm{\log}\left(\frac{M_{\rm BH}}{M_\odot}\right) = \log\left[ \left(\frac{\rm{FWHM}_{\rm H\beta}}{1000\,\rm{km\,s^{-1}}}\right)^2 \left(\frac{L_{5100}}{10^{44}\, \rm{erg\, s^{-1}}} \right)^{0.533}\right] + 7.03, 
\end{equation}

\noindent
where we assume, for concreteness, the virial factor for classical bulges.\footnote{Our main conclusions are qualitatively the same if we use the virial factor for pseudo bulges \citep{Ho2014}.}  The uncertainty of the BH masses is 0.35~dex. In Equation 3, ${\rm FWHM}_{\rm H\beta}$ gives the line width of broad H$\beta$ derived from the single-epoch spectra of \cite{Boroson1992}, and $L_{5100} \equiv \lambda L_\lambda (5100\, \rm{\mathring{A}})$, with $L_\lambda (5100\, \rm{\mathring{A}})$ the specific luminosity  of the nonstellar continuum at 5100 \AA.  

\begin{figure}[htbp]
\centering
\includegraphics[height=0.3\textheight]{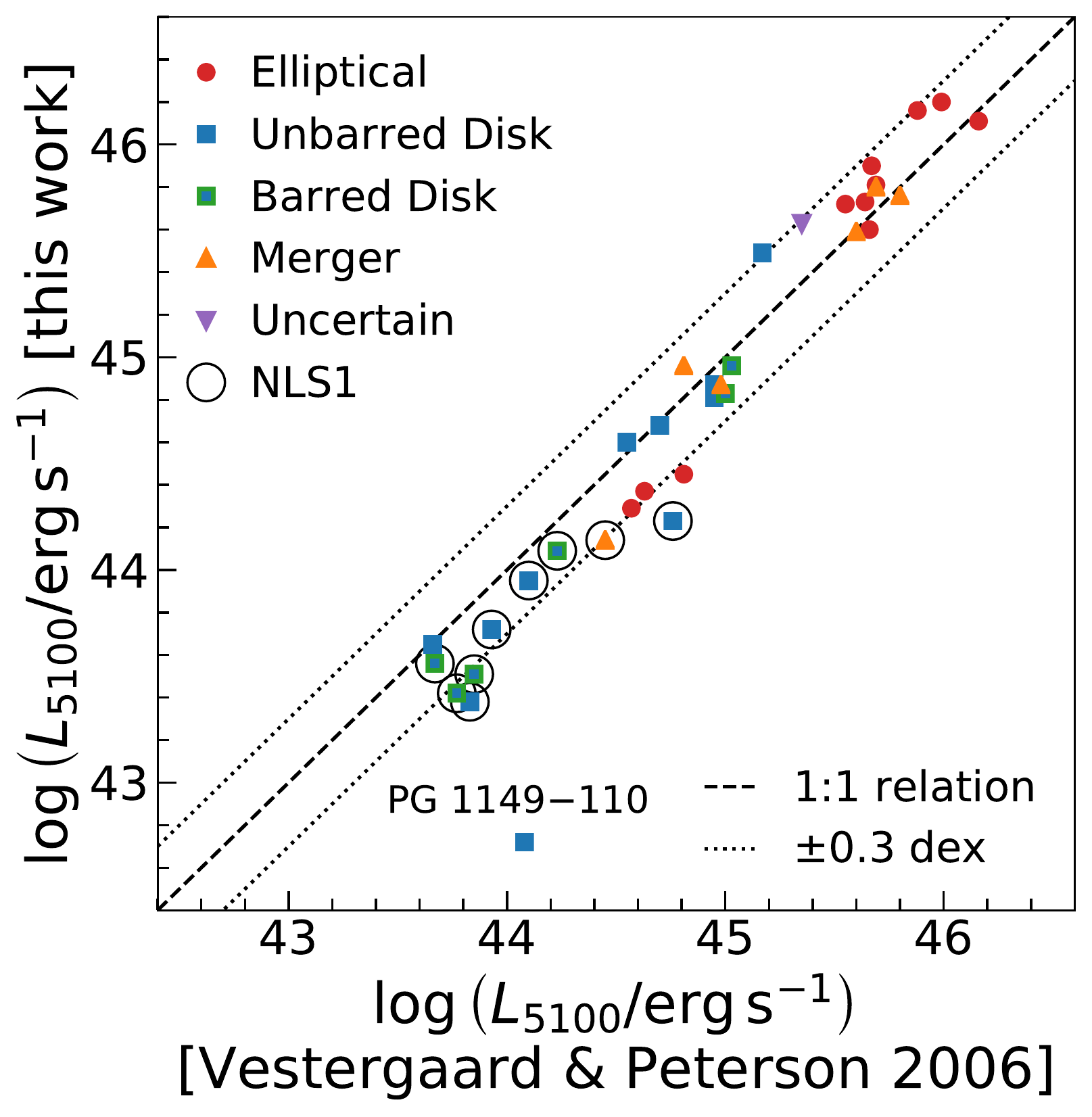}
\caption{The 5100~\AA\ continuum luminosity measured by our image decomposition is compared with the spectrophotometric measurements collected by \cite{Vestergaard2006ApJ}.  The dashed line is the 1:1 relation, with the dotted lines indicating $\pm 0.3$~dex.  The 0.05~dex uncertainties of \lagn\ are small and not plotted.  Our measurement of \lagn\ for PG~1149$-$110 is $\sim 20$ times lower than the previous measurement, likely because the source is variable (see text).
}
\label{Fig:l51}
\end{figure}

We obtain \lagn\ from our image analysis, converting the nuclear magnitude from the $B$-band \galfit\ decomposition to rest-frame 5100~\AA\ flux density assuming a power-law spectrum $f_{\nu} \propto \nu^{-0.44}$ \citep{Vanden_Berk2001} and correcting for Galactic extinction \citep{SF2011}.  Uncertainties arising from variations in PSF model and background subtraction (Section~\ref{sec:fit}) are typically less than 0.05 dex.  As shown in Figure~\ref{Fig:l51}, our values of \lagn\ agree within $\sim 0.3$~dex with those given by \cite{Vestergaard2006ApJ}, who made use of the ground-based spectrophotometry from \cite{Neugebauer1987ApJS}.  The lower luminosity sources with $L_{5100} \lesssim 10^{44.5}\,\rm{erg\, s^{-1}}$, predominantly narrow-line Seyfert 1s (NLS1s; $\mathrm{FWHM}_{\rm H\beta}\leq 2000\:\mathrm{km\:s^{-1}}$; \citealt{Osterbrock1985ApJ}), are systematically brighter in Neugebauer et al.'s measurements, no doubt because of the significant host galaxy light included in the large, $15\arcsec$-diameter aperture used for the ground-based observations.  The extreme outlying point for PG~1149$-$110 requires another explanation.  Its nucleus is quite modest in strength compared to the host in the $B$- and $I$-band HST images, both much fainter than reported in earlier observations (e.g., \citealt{Neugebauer1987ApJS,Jahnke2003MNRAS}).  Inspection of multi-epoch photometry in the W1 (3.4 $\mu$m) and W2 (4.6 $\mu$m) bands recorded during 2011--2019 by WISE \citep{Wright2010AJ} reveals that the light curves of PG~1149$-$110 reached a minimum in 2014, close in time to when our HST observations were conducted, and then quickly rose by more than 0.5 mag.  This level of near-IR variability, large even by the standards of ``changing-look'' AGNs, likely induces a brightnening of $\gtrsim 1$ mag in the optical (\citealt{Yang2018ApJ}).  Therefore, we suspect that for PG~1149$-$110 intrinsic variability is the culprit for the large difference between our measurement of its nuclear brightness and that recorded more than 20 years earlier.  The BH mass for this source was computed using  the value of \lagn\ from \cite{Vestergaard2006ApJ}.

\section{Discussion}
\label{Sec:Disc}

\begin{figure*}[htbp]
\centering
\includegraphics[height=0.3\textheight]{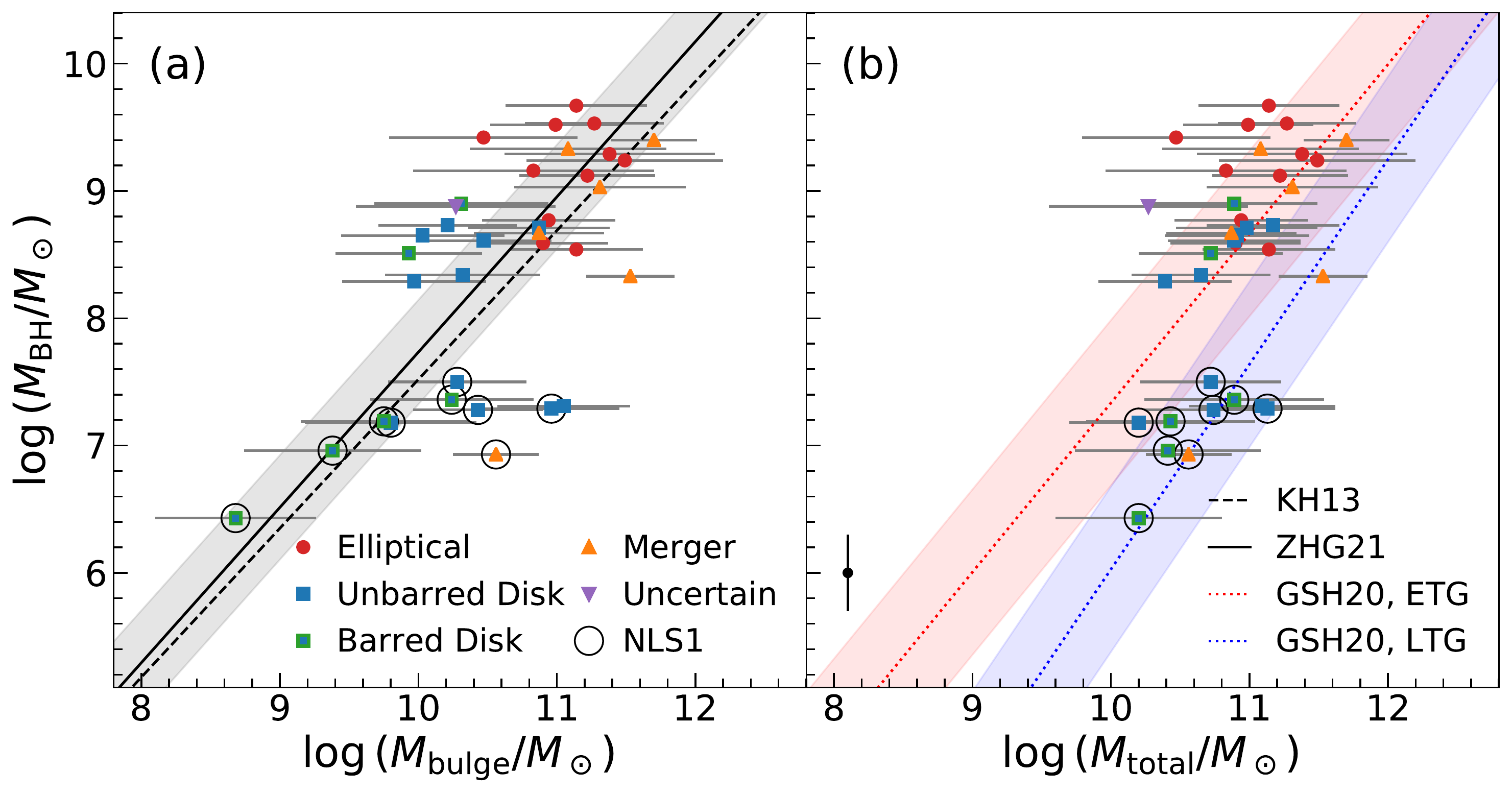}
\caption{The relation between BH mass and the stellar mass of (a) the bulge and (b) the entire host galaxy.  In panel (a), the dashed line is the $M_\mathrm{BH}-M_\mathrm{bulge}$ relation of classical bulges and elliptical galaxies from Kormendy \& Ho (2013; KH13), while the solid line is the revised  version from Zhu et al. (2021; ZHG21) that considers classical bulges and only the ``cores'' of elliptical galaxies.  In panel (b), the red and blue dotted lines are the $M_\mathrm{BH}-M_\mathrm{total}$ relation for early-type and late-type galaxies, respectively (Greene et al. 2020; GSH20).  The typical uncertainty of $M_\mathrm{BH}$ is given in the bottom-left corner.  The shaded regions indicate the intrinsic scatter of ZHG21 (0.41 dex), GSH20 ETG (0.65 dex), and GSH20 LTG (0.65 dex).  For claroty, the intrinsic scatter of KH13 (0.28 dex) is not displayed.}
\label{Fig:mbhgal}
\end{figure*}

\subsection{The \mbh--\mbulge\ and \mbh--\mhost\ Relations}
\label{sec:mbhgal}

Figure~\ref{Fig:mbhgal}a shows the relation between BH mass and bulge stellar mass for our sample of PG quasars.  For comparison, we overplot the \mbh--\mbulge\ relation of nearby, inactive elliptical galaxies and classical bulges from \cite{Kormendy2013}, as well as the revised version of this relation that replaces ellipticals by their ``core,'' the component most closely linked to the compact, high-redshift progenitors of present-day ellipticals, which were most directly related to the growth phase of the BH \citep{Zhu2021}.  As with \cite{Kormendy2013}, the zero point of the $M/L$ used by \cite{Zhu2021} was calibrated dynamically and does not depend on the choice of IMF.  Within the relatively large scatter of the quasar sample, both of the relations for inactive galaxies intersect our data points for the PG quasars, especially for the group BHs more massive than $M_{\rm BH} \gtrsim 10^{8}\,M_\odot$.  By contrast, the objects with $M_{\rm BH} \lesssim 10^{7.5}\,M_\odot$, which are predominantly NLS1s hosted by disk, and especially  barred, galaxies, depart notably from the $M_\mathrm{BH}-M_\mathrm{bulge}$ or $M_\mathrm{BH}-M_\mathrm{core}$ relations of their more massive counterparts.  This is qualitatively consistent with the known behavior of pseudo bulges in inactive galaxies \citep{Kormendy2013}, and suggests that the lower mass BHs reside in pseudo bulges.

While BH mass correlates most closely with bulge stellar velocity dispersion and stellar mass \citep{Kormendy2013}, its relation with the total galaxy stellar mass \citep{Reines2015}, albeit looser, nonetheless still offers a pragmatic and observationally less demanding empirical tool to investigate BH-galaxy coevolution.  Figure~\ref{Fig:mbhgal}b illustrates that the PG quasars obey the $M_\mathrm{BH}-M_\mathrm{total}$ relation of inactive galaxies, especially when separately divided into early-type galaxies and late-type galaxies, as recently calibrated by \cite{Greene2020}.  Note that \cite{Greene2020} derive stellar masses using the color-based $M/L$ from \cite{Bell2003}, which, on account of the ``diet Salpeter'' IMF assumed, need to be scaled by 0.06~dex to be consistent with our adopted \cite{Chabrier2003} IMF.

\begin{figure}[htbp]
\centering
\includegraphics[height=0.3\textheight]{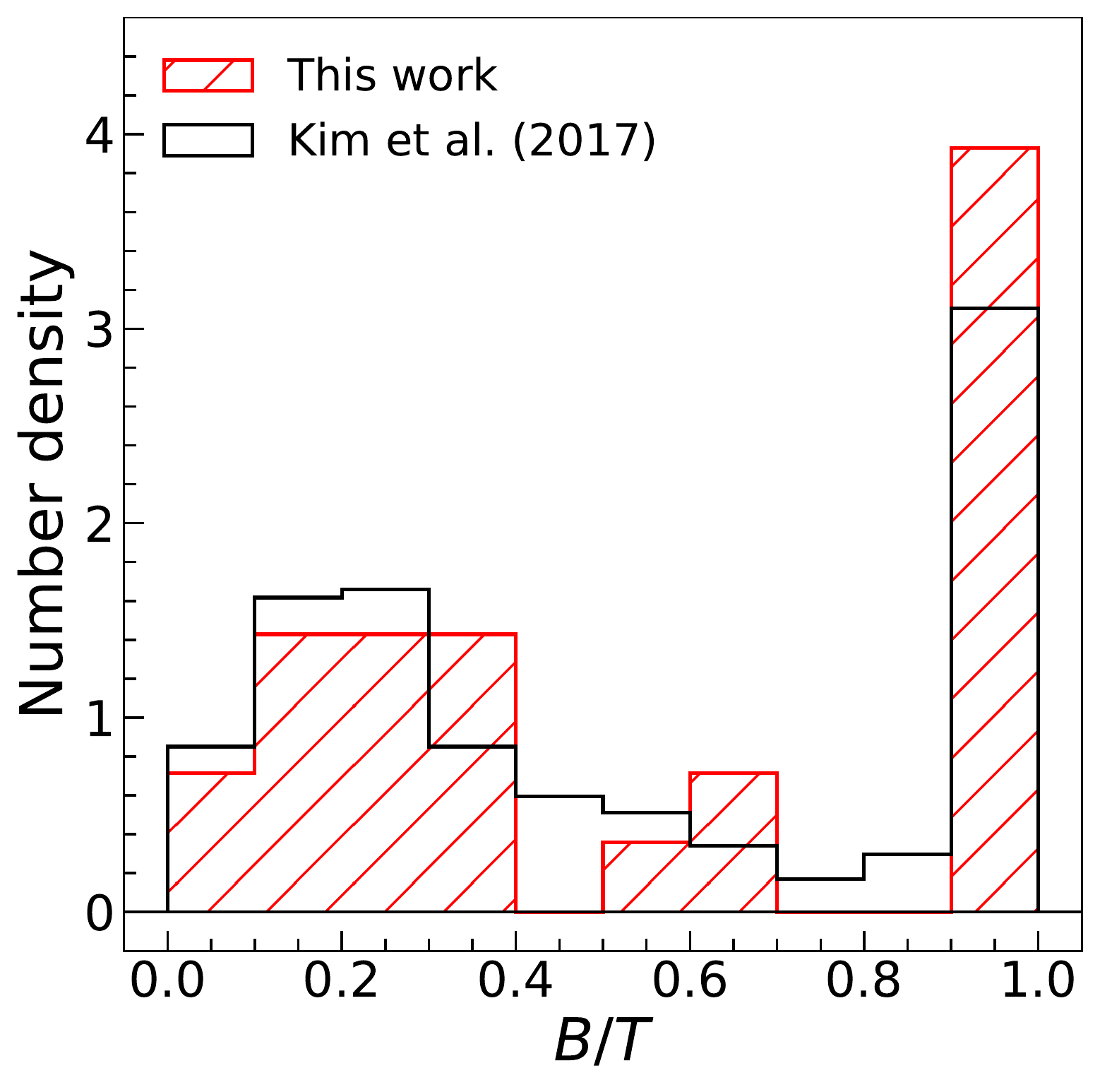}
\caption{Bulge-to-total ratio distribution of our sample (red hatched histogram) is in general consistent with that of the large sample of $z \lesssim 0.35$ AGNs from \cite[][black open histogram]{Kim2017}.}
\label{Fig:b2t}
\end{figure}

\begin{figure}[htbp]
\centering
\includegraphics[height=0.3\textheight]{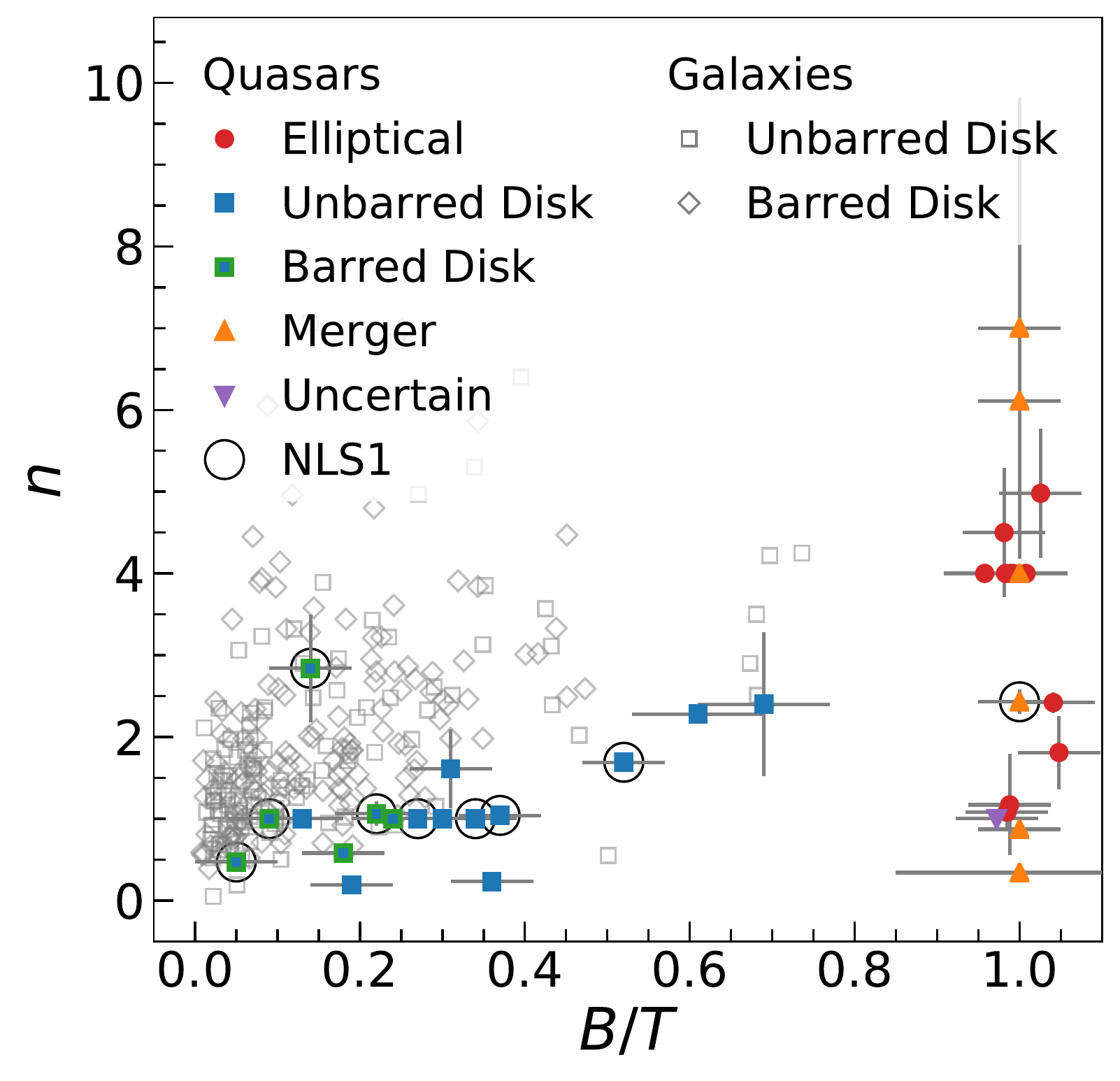}
\caption{The distribution of the bulge-to-total ratio ($B/T$) and bulge \sersic\ index ($n$) measured in the $I$ band.  Morphological types are given in the legend.  NLS1s are highlighted with circles.  Elliptical and systems with uncertain morphology are designated $B/T = 1$, as are mergers, for illustration purposes; for the sake of clarity, these points are moved slightly and randomly along the $x$ axis.  Shown for comparison in light grey symbols are the 320 disk galaxies with morphological decomposition from \cite{Gao2019}. }
\label{Fig:morph}
\end{figure}

\subsection{Diversity of Morphologies}

The most luminous quasars in our sample are mainly at $z \gtrsim 0.2$ and are thus not as well-resolved.  However, so long as they are detected robustly, their massive hosts can be securely classified as elliptical galaxies (Section~\ref{sec:morph}).  Moreover, these host galaxies usually display faint surface brightnesses and large effective radii that follow the Kormendy (1977) relation of inactive elliptical galaxies (Section~\ref{sec:kr}). Therefore, we conclude that most of the luminous ($M_B<-23.5$ mag or $\lagn \gtrsim 10^{45}\,\mathrm{erg\,s^{-1}}$) quasars in our sample likely reside in elliptical galaxies, consistent with early HST observations \citep{Disney1995Natur,Bahcall1997ApJ,McLure1999MNRAS,Dunlop2003}.

Figure~\ref{Fig:b2t} compares the distribution of bulge-to-total ratio of our sample with that of the 235 type~1 AGNs studied by \cite{Kim2017}, which, while heterogeneous, is the largest and most representative sample to date of nearby ($z\lesssim 0.35$) AGNs with detailed morphological decomposition based on HST imaging.  The Anderson-Darling test (Scholz \& Stephens 1987) cannot reject the null hypothesis that the samples are drawn from the same parent distribution ($p$-value $>0.25$).  Our sample contains a moderately higher fraction of elliptical ($B/T=1$) galaxies, likely because our sample extends to slightly higher redshift ($\sim 0.4$) than that of \cite{Kim2017}, and it comprises among the most luminous sources within its redshift range.  As in \cite{Kim2017}, mergers make up only a minority (17\%; 6/35) of the sample, emphasizing that mergers are not a prerequisite for triggering quasar-level activity in galaxies (\citealt{Villforth2014MNRAS,Hewlett2017MNRAS}). 

Quasars hosted by disk galaxies exhibit relatively modest bulge fractions ($B/T \lesssim 0.5$) and low bulge \sersic\ indices ($n \lesssim 2$).  Those that are barred, in particular, have the lowest $B/T$ (Figure~\ref{Fig:morph}), consistent with the overall trend observed among nearby inactive galaxies (gray symbols) from the Carnegie-Irvine Galaxy Survey \citep{Ho2011} with detailed morphological decomposition \citep{Gao2019}.\footnote{Although the analysis of \cite{Gao2019} was based on the $R$ band, differences betweem the $R$ and $I$ band should not be large (e.g., \citealt{Tasca2011AA}).} In contrast, the elliptical galaxies and major mergers have large \sersic\ indices.  While rigorous distinction between pseudo and classical bulges is still controversial (\citealt{Gao2020} and references therein), pseudo bulges are generally characterized by low $B/T$ and low \sersic\ $n$.  Adopting, for simplicity, $n<2$ as a criterion for recognizing pseudo bulges, 82\% (14/17) of the PG quasars hosted by disk galaxies contain a pseudo bulge.  

Our sample contains a total of nine NLS1s (Figure~\ref{Fig:morph}). All but one reside in disk galaxies, among which 50\% (4/8) are barred.  This fraction is consistent with that of \cite{Kim2017} but lower than that of earlier studies of (lower luminosity) Seyfert galaxies (e.g., 65\%--85\%; \citealt{Crenshaw2003AJ,Ohta2007}).  The bar fraction of inactive disk galaxies decreases with increasing galaxy stellar mass \citep{Diaz2016,Li2017,Erwin2018,Zhao2020}, averaging $\sim$50\% for stellar masses $\sim 10^{10}-10^{11}\,M_\odot$, similar to our finding.  The higher bar fraction reported in previous NLS1 samples likely can be attributed to their lower typical galaxy masses.  The lower bar fraction of 22\% (2/9) among the targets with ${\rm FWHM}_{\rm H\beta} > 2000\,\mathrm{km\:s^{-1}}$ (``broad-line'' objects), while in agreement with previous studies \citep{Crenshaw2003AJ,Ohta2007,Kim2017}, nevertheless may be underestimated by observational bias \citep{Erwin2018} due to the larger distances of the sample (median $z \approx 0.14$ compared to $z \approx 0.05$ for the NLS1s). 

With $R \equiv L_\nu({\rm 6~cm})/L_\nu(B) > 100$, six of the quasars in our sample (Table~\ref{tab:basic}) can be considered radio-loud (Kellermann et al. 1989). Apart from a single exception in a merger, the rest of the radio-loud sources are hosted by massive ellipticals, which are systematically more massive than their radio-quiet counterparts.  \cite{Dunlop2003} and \cite{Kim2017} reached similar conclusions based on more extensive statistics.

\begin{figure}[htbp]
\centering
\includegraphics[height=0.3\textheight]{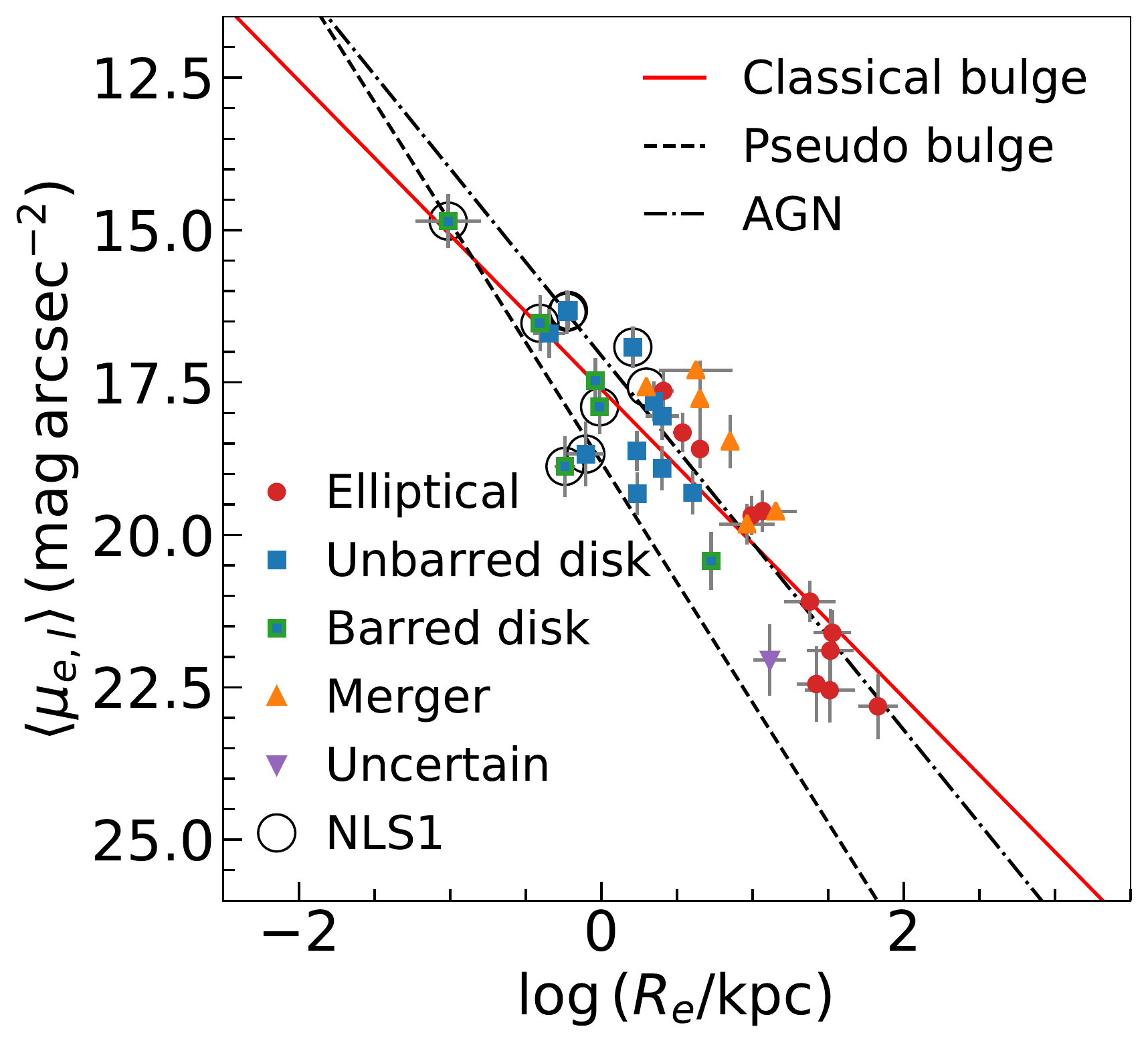}
\caption{The Kormendy relation between effective radius ($R_e$) and mean surface brightness within $R_e$ (\mmue) for the bulges of AGN host galaxies.  The best-fit relations from \cite{Kim2019} are shown for classical bulges and elliptical galaxies (solid red line), pseudo bulges (dashed line), and AGNs (dash-dotted line).  We converted the relations of \cite{Kim2019} from the $R$ band to the $I$ band assuming $R-I=0.7$ mag \citep{Fukugita1995PASP}.  
}
\label{Fig:kr}
\end{figure}

\subsection{Kormendy Relation}
\label{sec:kr}

The effective radius ($R_e$) forms a strong inverse correlation with the mean surface brightness within $R_e$ in elliptical galaxies \citep{Kormendy1977ApJ} and classical bulges \citep{Carollo1999ApJ}.  By contrast, pseudo bulges depart markedly from the Kormendy relation by manifesting lower $\langle \mu_{e} \rangle$ at a given $R_e$ \citep{Kormendy2004,Gadotti2009MNRAS,Fisher2010ApJ,Gao2020}.  Early HST observations revealed that quasar host galaxies, when properly decomposed from the bright nucleus, follow the Kormendy relation of inactive galaxies \citep{Dunlop2003}.  However, with the vastly expanded and more diverse sample of type~1 AGNs of \cite{Kim2017}, \cite{Kim2019} found that the bulges of AGN host galaxies tend to be more luminous than those of inactive galaxies of the same bulge size, concluding that their bulge surface brightnesses are likely enhanced by newly formed stars associated with the current episode of BH accretion.

We calculate the mean effective surface brightness by averaging the flux inside the effective radius of the bulge,
 
\begin{equation}\label{eq:mu}
\langle \mu_{e,I} \rangle = M_I + 5\log\left(\frac{R_e}{10\,\mathrm{pc}}\right) + 28.57,
\end{equation}

\noindent
where $M_I$ is absolute magnitude of the bulge.  As shown in Figure~\ref{Fig:kr}, the bulges of PG~quasars generally overlap with the distributions defined by the classical bulges of inactive galaxies and AGNs studied by \cite{Kim2019}.  At the same time, the quasar hosts designated by us as ellipticals in this study indeed possess large effective radii and correspondingly faint surface brightnesses similar to their inactive counterparts \citep{Kormendy2009,Huang2013,Gao2020}, which lends confidence in our morphological assignment.  Major mergers tend to have systematically higher surface brightnesses than the rest of the sample, likely a reflection of boosted central brightness from merger-induced starbursts, by analogy with ultra-luminous infrared galaxies \citep{Rothberg2013ApJ}.  At a given $R_e$, pseudo bulges in inactive galaxies display lower surface brightness than classical bulges or elliptical galaxies \citep{Kormendy2004,Gadotti2009MNRAS,Fisher2010ApJ,Gao2020}.  Judging by a variety of evidence (disk morphologies, presence of a bar, low $B/T$, low $n$), a number of the PG quasars in our sample, particularly those classified as NLS1s, should possess pseudo bulges, but, intriguingly, they are located mostly {\it above}\ the Kormendy relation of pseudo bulges.  As for the mergers, the simplest explanation for this trend is to posit that the AGNs hosted by pseudo bulges have central surface brightnesses enhanced by recent or ongoing star formation.  \cite{Zhao2019} reached the same conclusion studying the HST images of nearby type~2 quasars.  

The results presented here, in concert with the analysis of \cite{Kim2019}, paint a consistent picture. The host galaxies of quasars, most notably those with accretion rates high enough to be NLS1s, which tend to be relatively late-type, often barred, disk galaxies with pseudo bulges, show evidence of enhanced young stellar populations relative to their inactive counterparts.  Mergers do as well---whether they be active or not---when compared to classical bulges and ellipticals.  Nearby AGNs, notably those powerful enough to qualify as quasars, lack neither gas nor the ability to form stars (e.g., \citealt{Husemann2017MNRAS,Shangguan2018,Shangguan2020ApJS, Jarvis2020,Yesuf2020,Zhuang2020ApJ,Xie2021}). BH accretion and host galaxy star formation are tightly linked by the molecular gas \citep{Shangguan2020ApJ,Zhuang2021}, which may be preferentially more centrally concentrated \citep{Molina2021}.

\section{Conclusions}
\label{Sec:Con}

We investigate the host galaxies of 35 low-redshift ($z<0.5$) Palomar-Green (PG) quasars with high-resolution HST/WFC3 images in the rest-frame $B$ and $I$ bands.  With the aid of multi-component 2D image decomposition, we classify the morphological types of the host galaxies, measure the structural parameters of their bulge, constrain the stellar population, and estimate the stellar mass and its relation to the central BH.  Our main results are as follow. 
 
\begin{itemize}
\item About half of our quasar host galaxies possess a disk, and hence morphologically can be classified as spiral or S0 galaxies, among which $\sim 1/3$ are barred.  Most of the lower luminosity quasars are NLS1s, and they tend to reside in barred disks galaxies with pseudo bulges.  The more powerful quasars ($M_B < -23.5$ mag or $\lagn \gtrsim 10^{45}\,\mathrm{erg\,s^{-1}}$) are hosted by elliptical galaxies.  Mergers account for less than 20\% of the sample, with no apparent connection to the radio-loudness of the AGN.  

\item The host galaxy-decomposed $B$-band magnitude of the nucleus agrees well with early spectrophotometric measurements.  We use the new optical nuclear magnitudes to update the single-epoch virial BH masses. 

\item Using the $B-I$ color to constrain the stellar mass, we find that PG quasars generally track the scaling relations between BH mass and stellar mass observed in local inactive galaxies.  BHs more massive than $\sim 10^8\,M_\odot$ follow the \mbh--\mbulge\ and \mbh--$M_{\rm core}$ relations of classical bulges and ellipticals, as well as the
\mbh--\mhost\ relation of early-type galaxies; less massive BHs, particular those in NLS1s, behave like pseudo bulges and late-type galaxies.  The correlations between BH mass and bulge stellar mass or total galaxy stellar mass 
are well established in low-$z$ quasar host galaxies.

\item The bulges of the host galaxies of PG quasars display a relatively tight correlation between effective radius and mean surface brightness, one that coincides with the Kormendy relation of inactive elliptical galaxies and classical bulges, in spite of the fact that $\sim 50\%$ of the disk and barred galaxies likely contain pseudo bulges.  In agreement with findings based on other samples of nearby quasars, we argue that a sizable fraction of the disk host galaxies, in particular those that possess pseudo bulges, exhibit central regions brightened by recent or ongoing star formation.
\end{itemize}

\acknowledgements
We thank the anonymous referee for useful suggestions to improve the clarity of the paper.  This research was supported by the National Science Foundation of China (11721303 and 11991052) and the National Key R\&D Program of China (2016YFA0400702).  DZ was supported by the Scholar Program of Beijing Academy of Science and Technology (BS202002).  MK was supported by the National Research Foundation of Korea (NRF) grant funded by the Korea government (MSIT) (No. 2020R1A2C4001753). YZ thanks Yunfeng Chen and Niankun Yu for advice on data analysis.  JS thanks Qian Yang for discussion of quasar variability.  

\facility {HST}	

\clearpage
\appendix

\section{WCS Correction for {\tt GYRO}-mode Observations}
\label{Append:WCS}

HST pointing is fine-locked when there are two guide stars available during the exposure, in which case the pointing can reach an accuracy of $\sim 2-5$ mas during an orbit.  However, when only one guide star is available and gyros are used to control the telescope in {\tt GYRO} mode, the exposure can incur a typical drift rate of 1--2 $\mathrm{mas\,s^{-1}}$ ($\sim 5\, \mathrm{mas\,s^{-1}}$ in rare cases; \citealt{Gonzaga2012}), and the final image will be smeared if we combine the dithered frames without correcting the world coordinate system (WCS).

Four of our targets were observed in {\tt GYRO} mode (Table \ref{tab_obj}).  For the typical integration times in our program, the WCS mismatch in {\tt GYRO} mode is up to $\sim 1\farcs4$.  While the drift during the exposure of each dither cannot be removed, the relative drift between the individual dithered frames can be corrected.  The standard method to correct WCS offset is to align the images with {\tt TweakReg} \citep{Gonzaga2012} using reference sources in the field.  However, the small field-of-view of our observations and their relatively shallow depth conspired to yield very few reference sources in the images suitable to for proper coordinate registration.  This compelled us to align the dithered images manually.  We use the bright nucleus to adjust the WCS reference in the image header.  For the UVIS observations, we use the unsaturated short exposure as the reference image, and the first sub-exposure is used as the reference image for the IR observations.  We subtract the other sub-exposures from the reference image, iteratively fine-tuning their WCS to minimize the asymmetric residuals that result from astrometric misalignment.  We estimate that this method can reach an aligment accuracy of $\sim 1.5 - 3\,\mathrm{mas}$.  Figure~A1 shows the  drizzle-combined images of the nucleus of PG~0050+124 without and with manual WCS correction.

\begin{figure}[htbp]
\begin{center}
\figurenum{A1}
\includegraphics[width=0.5\textwidth]{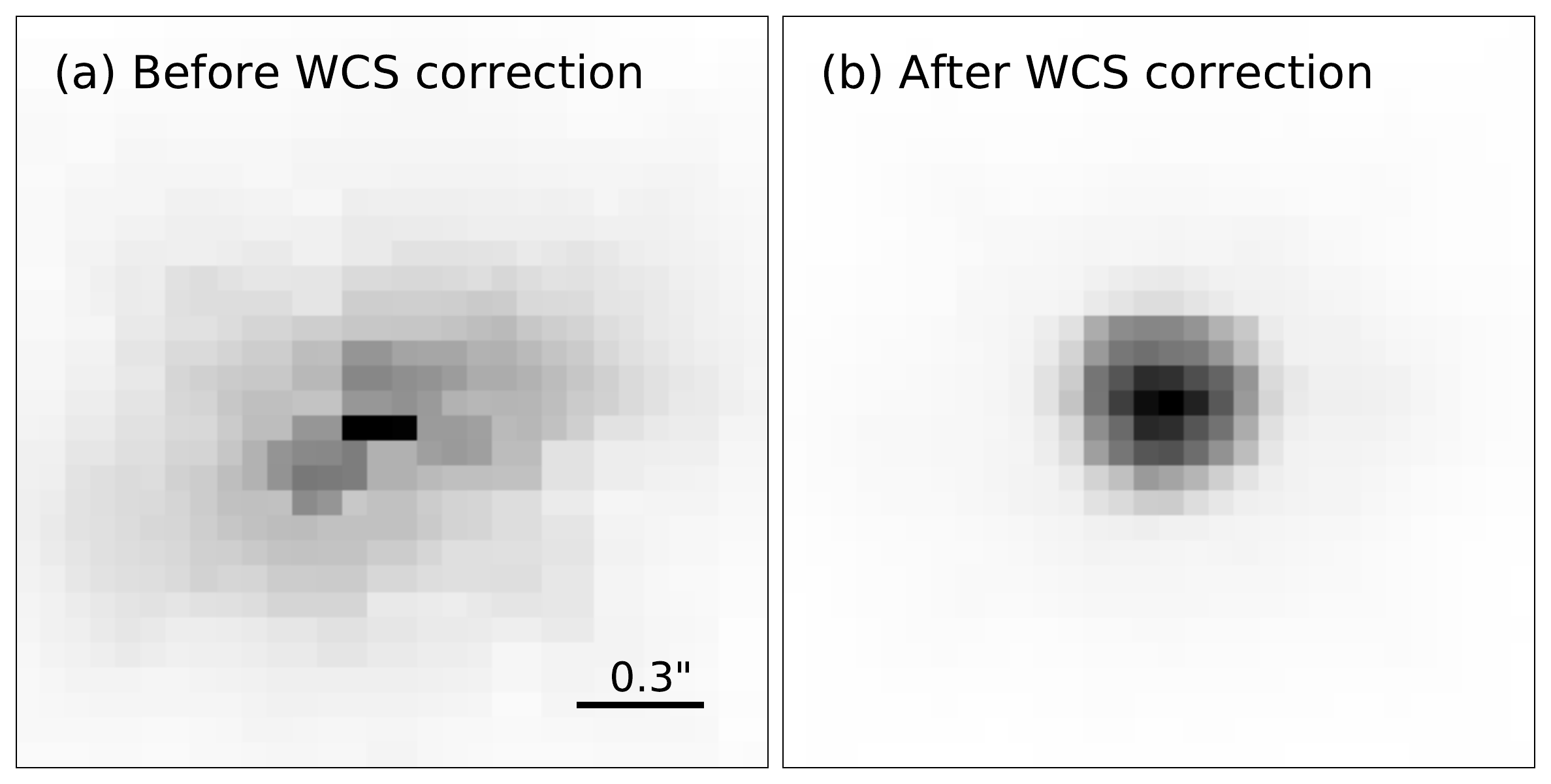}
\caption{The drizzle-combined WFC3/F105W image of the nucleus of PG~0050+124 (a) before and (b) after manual WCS correction.}
\label{Fig:WCS_cor}
\end{center}
\end{figure}

\section{Notes on Individual Objects}
\label{apd:note}

PG~0050+124: The host features a grand-design spiral, with one of the arms mildly perturbed by a small, nearby companion, which as masked during the fit. We model the galaxy with a \sersic\ bulge (free $n$) and an exponential ($n=1$) disk.  A weak, bar-like feature may be present, but adding a bar component does not significantly improve the residuals, and therefore we do not include it.  

PG~0052+251: A bar component is clearly present, but to account for it the \sersic\ index of the bulge component has to be fixed. The smallest residuals are achieved using a bulge with $n=1$.  The best-fit model leaves a residual outer ring structure, which we do not attempt to fit.

PG~0804+761: The host has an obvious extended disk with a bar along the direction of the diffraction spike. We fit it with three components, one for the bulge (free $n$), one for the disk ($n=1$), and one for the bar ($n=0.5$).

PG~0923+129: The host has an obvious extended disk with a very diffuse outer ring.  We fit it with three free \sersic\ components, one for a bulge, one for a disk, and a third for an outer ring.  Spiral structure can be seen in the central $\sim 3\arcsec$, but it is too faint to be properly fit.

PG~0934+013: The host galaxy contains an obvious disk with a prominent bar and spiral arms. The bulge is fit with $n$ set free.  The bulge component in the F438W filter is very faint and difficult to fit; its magnitude is estimated by iteratively adjusting the amplitude of the F814W model until the residuals appear acceptable.  The $B$-band magnitude of the bulge is likely more uncertain than in other sources, although the final $B-I$ color ($2.20\pm0.70$ mag) does not seem unusual.

PG~0953+414: The host galaxy is very faint compared to the nucleus.  We can obtain a useful fit only with a single-component model with a fixed \sersic\ index.  The best fit favors $n=1$. We deem the morphology too uncertain to be classified.

PG~1011$-$040: The host galaxy contains an obvious disk with a prominent bar and spiral arms.  The bulge is fit with $n$ set free.

PG~1012+008: This is an obvious merging system of two, near-equal brightness galaxies, with another possible small companion projected to the north.  The host galaxy of the AGN can be fit with two \sersic\ components modified by an $m = 1$ Fourier mode.  A similar model is used for the main interacting companion.  The small northern companion is fit simultaneously with a single \sersic\ component.  We masked the northern companion when calculating the nonparametric magnitude of the system.

PG~1022+519: The host galaxy can be well fit with a bulge component (free $n$), bar, and disk modified by an $m=1$ Fourier mode.

PG~1049$-$005: The host galaxy shows obvious tidally-disrupted feature. We use two \sersic\ components modified by an $m = 1$ Fourier mode to fit the host and an additional component to account for the north tidal tail structure.
 
PG~1100+772:  We fit the galaxy with a single component.  The \sersic\ index is somewhat low ($n\approx 1.1$), and there may be mild evidence for additional structure.  

PG~1103$-$006: We fit the host galaxy with a single \sersic\ model with $n$ set free.  A faint projected companion to the west of the quasar is fit simultaneously. Other sources in the field are masked.

PG~1114+445: The galaxy is fit with two free \sersic\ components for the bulge and disk.  An outer ring-like structure is not considered.

PG~1119+120: We adopt a bulge with $n = 1$ and an exponential disk, plus an additional \sersic\ component to account for the faint companion to the north.  We tried to add a bar, but it did not improve the fit.  

PG~1149$-$110: The host galaxy is fit with a bulge (free $n$) and disk.  There is a faint, diffuse ring in the outskirts of the galaxy.

PG~1202+281: The host galaxy can be well fit by a single \sersic\ component.  The projected companion is modeled with another \sersic\ component.

PG~1216+069: The nuclear emission is extremely dominant.  We fit the host with a single \sersic\ component, masking the nearby, highly saturated star, which cannot be fit simultaneously.  

PG~1226+023: The nuclear emission is extremely dominant.  We fit the host galaxy with a single \sersic\ component, fixing $n$ to achieve a stable solution. Both $n=3$ and $n=4$ yield acceptable fits, and we adopt $n=4$ model as the final model, with the difference between $n=3$ and $n=4$ models bracketing the flux uncertainty. The well-known jet is seen in the residuals.

PG~1244+026: The host galaxy can be well fit with a bulge, bar, and disk.  Fixing the bulge \sersic\ index to 1 gives the lowest residuals.  The disk contains faint spiral structure, but we do not attempt to model it.  The nucleus is slightly elongated likely due to the pointing drift (Appendix A).  

PG~1259+593: The nuclear emission is very bright.  We fit the galaxy with only a single \sersic\ component with $n$ fixed to 4, even though there is mild evidence for additional faint, extended emission in the residuals.

PG~1302$-$102:  There is a plume of extended emission to the north that may be remnant of a merger.  Two small, compact sources are projected toward the center of the galaxy.  Apart from accounting for the projected sources, we fit the quasar host with two \sersic\ components including Fourier modes.  

PG~1322+659: This is an obvious spiral galaxy.  The bulge was fit with a \sersic\ component with $n=1$, and the spiral arms were modeled with coordinate rotation and Fourier modes.

PG~1351+640: The host galaxy can be well fit with a single \sersic\ component.
	
PG~1354+213: The host galaxy requires two components, which together can be interpreted to be a bulge plus disk system.  However, the bulge has a very small \sersic\ index ($n\approx 0.2$); fixing it to a larger value produces notably worse residuals.

PG~1444+407: The host galaxy requires two components, which together can be interpreted to be a bulge plus disk system. However, the bulge has an obviously elongated morphology and a \sersic\ index suspiciously close to that of a bar ($n\approx 0.2$).  For the sake of concreteness, we attribute it to the bulge.

PG~1448+273: The host galaxy is clearly a late-stage merger.  We adopt three \sersic\ components to fit it, but there remains substantial structure in  the residual image.

PG~1534+580: The host galaxy can be well fit with a bulge (free $n$) and an exponential disk.  Very faint spiral features are seen in the residuals.

PG~1535+547: The host galaxy is fit with an $n=1$ bulge and an exponential disk.  Allowing the \sersic\ index to be free yields a best fit value of 
$n\approx 0.5$, but the residuals are similar to the case of $n=1$, and we finally adopt the latter.

PG~1613+658: The host galaxy is a dramatic late-stage merger (see also Hong et al. 2015), with a projected compact source that may be the second nucleus. We adopt two free \sersic\ components to fit the main body of the host and another component to account for the projected companion.  The distant galaxy to the north is masked.

PG~1626+554: The host galaxy can be well fit with an $n=1$ bulge and an exponential disk.  Allowing the \sersic\ index to be free yields a best fit value of $n\approx 0.2$, but the residuals are similar to the case of $n=1$, and we finally adopt the latter.  Faint spirals are evident in the residuals.  

PG~1700+518: The nuclear emission is very bright. We fit the host, which appears to be a late-stage merger, with a single \sersic\ component fixed to $n=4$ plus and additional component to account for the arc-like feature to the north.

PG~1704+608: The nuclear emission is very bright.  We adopt a single \sersic\ component ($n=4$) to fit the host galaxy.

PG~2112+059: The nuclear emission is very bright.  We adopt a single \sersic\ component ($n=4$) to fit the host galaxy.

PG~2214+139: The host galaxy can be well fitted with a single \sersic\ component.  There are faint, ring-like features in the residuals, which are related to the shells seen in the deeper ground-based image of \cite{Hong2015}.

PG~2251+113: The nuclear emission is very bright.  We adopt a single \sersic\ component ($n=4$) to fit the host galaxy.

\begin{longrotatetable}
\begin{deluxetable}{lccccccccccrc}
\label{tab_obj}
\tablewidth{0pt} 
\tablecaption{Object Information \label{tab:basic}}
\tablehead{ 
\colhead{Name} & 
\colhead{Alternative name} &
\colhead{Observation} &
\colhead{Filter} & 
\colhead{Exposure Time} & 
\colhead{$z$} & 
\colhead{$D_{L}$}  & 
\colhead{$A_{V}$}& 
\colhead{$M_V$} &
\colhead{$\log\,R$} &
\colhead{$\log\,L_{5100}$} & 
\colhead{$\mathrm{FWHM_{H\beta}}$} & 
\colhead{$\log\,M_\mathrm{BH}$} \\
 & 
 & 
\colhead{Date} &
($B$/$I$)& 
\colhead{(s)} &
 &
\colhead{(Mpc)} & 
\colhead{(mag)} & 
\colhead{(mag)} & 
 &
\colhead{($\mathrm{erg\,s^{-1}}$)} & 
\colhead{($\mathrm{km\,s^{-1}}$)} & 
\colhead{($M_\odot$)}
}
\colnumbers
\startdata
PG 0050+124\tnm{*}  & I Zw 1       & 2013-11-02 & F438W/F105W & 300/147 & 0.061 &   282.1 & 0.171 & -23.77 & $-0.48$ & 44.23 & 1171 & 7.29  \\ 
PG 0052+251         &              & 2013-07-31 & F475W/F105W & 360/147 & 0.155 &   762.5 & 0.125 & -24.64 & $-0.62$ & 44.83 & 5187 & 8.90  \\ 
PG 0804+761         &              & 2013-03-10 & F438W/F105W & 450/147 & 0.100 &   475.0 & 0.095 & -24.44 & $-0.22$ & 44.96 & 3045 & 8.51  \\ 
PG 0923+129         & Mrk 705      & 2012-12-16 & F438W/F814W & 300/180 & 0.029 &   131.1 & 0.109 & -21.59 & $ 0.32$ & 43.38 & 1957 & 7.28  \\ 
PG 0934+013         & Mrk 707      & 2013-01-29 & F438W/F814W & 450/300 & 0.050 &   229.4 & 0.184 & -21.43 & $-0.42$ & 43.51 & 1254 & 6.96  \\ 
PG 0953+414         &              & 2013-02-03 & F475W/F105W & 300/147 & 0.239 &  1234.6 & 0.032 & -25.65 & $-0.36$ & 45.63 & 3111 & 8.88  \\ 
PG 1011$-$040       &              & 2013-03-29 & F438W/F814W & 300/210 & 0.058 &   267.7 & 0.097 & -22.70 & $-1.00$ & 44.09 & 1381 & 7.36  \\ 
PG 1012+008         &              & 2013-05-03 & F475W/F105W & 360/147 & 0.185 &   926.7 & 0.093 & -24.79 & $-0.30$ & 44.87 & 2615 & 8.33  \\
\enddata
\tablecomments{ 
Column (1): Object name.
Column (2): Alternative name of the object.
Column (3): Date of observations.
Column (4): Filters for rest-frame $B$ and $I$ band.
Column (5): Total long exposure time for rest-frame $B$ and $I$ band image.  
Column (6): Redshift from \cite{Boroson1992}.
Column (7): Luminosity distance.
Column (8): Galactic extinction in the $V$ band, estimated from \cite{SF2011}.                           
Column (9): $V$-band absolute magnitude from \cite{Boroson1992}.
Column (10): Ratio of radio-to-optical luminosity from \cite{Kellermann1989}.
Column (11): AGN 5100~\AA\ continuum luminosity converted from $B$-band nuclear emission derived from our GALFIT decomposition.
Column (12): FWHM of broad H$\beta$ line from \cite{Boroson1992}.
Column (13) BH mass calculated in Section~\ref{sec:mbh}.
}
\tablenotetext{*}{Observed in GYRO mode.}
\end{deluxetable}%
                                                                                                                                              
\end{longrotatetable}

\end{document}